\documentclass[twocolumn,tighten]{aastex6}

\usepackage{graphicx}	
\usepackage{amsmath}	
\usepackage{amssymb}	
\usepackage{upgreek}

\newcommand\xmm{{\it XMM-Newton}}
\newcommand\suzaku{{\it Suzaku}}
\newcommand\nustar{{\it NuSTAR}}
\newcommand\swift{{\it Swift}}

\newcommand\ergsec{{\rm~erg\ s}^{-1}}

\newcommand\mpc{{\rm~Mpc}}

\begin{document}

\title{Long-term Spectral Variability of the Ultra-luminous X-ray source Holmberg IX X--1}
\shorttitle{Variability of ULX Ho IX X--1}

\author{V. Jithesh$^1$, Ranjeev Misra$^2$ and Zhongxiang Wang$^1$}
\affil{$^{1}$ Shanghai Astronomical Observatory, Chinese Academy of Sciences, 80 Nandan Road, Shanghai 200030, China; jithesh@shao.ac.cn \\
$^{2}$ Inter-University Centre for Astronomy and Astrophysics, Post Bag 4, Ganeshkhind, Pune 411007, India}
\shortauthors{V. Jithesh et al.}

\begin{abstract}
We investigate the long-term spectral variability in the ultra-luminous 
X-ray source Holmberg IX X--1. By analyzing the data from eight {\it Suzaku} 
and 13 {\it XMM-Newton} observations conducted between 2001 and 2015, we 
perform a detailed spectral modeling for all spectra with simple models 
and complex physical models. We find that the spectra can be well explained 
by a disc plus thermal Comptonization model. Applying this model, we unveil 
correlations between the X-ray luminosity ($L_{\rm X}$) and the spectral parameters. 
Among the correlations, a particular one is the statistically 
significant positive correlation between $L_{\rm X}$ 
and the photon index ($\Gamma$), while at the high luminosities of 
$> 2\times10^{40}\,\ergsec$, the source becomes marginally hard and that 
results a change in the slope of the $\Gamma - L_{\rm X}$ correlation. 
Similar variability behavior is observed in the optical depth of the source 
around $L_{\rm X} \sim 2\times10^{40}\,\ergsec$ 
as the source becomes more optically thick. We consider the scenario that a 
corona covers the inner part of the disc, and the correlations 
can be explained as to be driven by the variability of seed photons from the 
disc input into the corona. On the basis of the disc-corona model, we discuss 
the physical processes that are possibly indicated by the variability of the 
spectral parameters. Our analysis reveals the complex variability behavior 
of Holmberg IX X--1 and the variability mechanism is likely related to the 
geometry of the X-ray emitting regions.

\end{abstract}


\section{Introduction} 
\label{sec:intro}

Ultra-luminous X-ray sources (ULXs) are extragalactic, point-like, non-nuclear 
X-ray sources with observed X-ray luminosity, $L_{\rm X} > 10^{39} \ergsec$, 
exceeding the Eddington limit for a typical stellar-mass 
($\sim 10\,\rm M_{\odot}$) black hole \citep[see][for a review on ULXs]{Fen11}. 
Studies with {\it Advanced Satellite for Cosmology and Astrophysics} ({\it ASCA}) 
observations \citep{Mak00} revealed the X-ray flux variability on different 
time-scales, which suggests the binary accreting nature for ULXs. Early 
spectra of ULXs were described by a multi-color disc blackbody \citep[MCD;][]{Mit84} 
plus power law (PL) model, mimicking the popular spectral model for Galactic 
black hole X-ray binaries \citep[BHXBs; See][]{Rem06}. Such modeling has 
provided evidence of cool thermal components \citep[$kT_{\rm in} 
\sim 0.20$ keV;][]{Mil03}, which has been interpreted as the existence of 
intermediate-mass black holes (IMBHs) of mass $\sim 10^{2} - 10^{4}\,\rm M_{\odot}$ 
in ULXs \citep{Col99, Mil04a}. However, this interpretation has been challenged by the highest 
quality \xmm{} observations, where the observed spectra show a broad curvature 
\citep{Sto06} at high energies ($\gtrsim 3$ keV), which does not correspond 
to any of the known sub-Eddington accretion states in BHXBs \citep{Rem06} and 
is hardly reconciled with the IMBH interpretation \citep{Rob07}. The peculiar 
features (a soft component and the curvature at the high energies) in the highest 
quality data suggest a new observational state for ULXs, which is referred
as {\it ultraluminous state} \citep{Gla09}. Moreover, the observed 
properties of the ULXs are better explained on basis of disc plus 
Comptonized corona models in this state and such modeling suggests 
that the majority of ULXs are black holes of stellar origin accreting at 
near-Eddington and/or super-Eddington rate \citep{Sto06, Pin12, Sut13}. 
In addition, the recent discovery of pulsating neutron stars (NSs) in 
the three ULXs, M82 X--2 \citep{Bac14}, NGC 5907 X--1 and NGC 7793 P13 \citep{Isr17a, Isr17b},
has proved the existence of NSs as the compact primaries in 
the ULX population. It is also noted that typical 
ULX spectra can be described equally well with phenomenological models adopted 
for Galactic X-ray pulsars \citep{Pin17}. Thus a few 
non-pulsating ULXs may host NSs. 

Spectral variability has been studied for individual as well as a sample 
of ULXs \citep{Fen06, Fen09, Kaj09, Pin12}. In individual ULXs, the 
monitoring observations have been used to trace the long-term variability, 
which sometimes was interpreted as state transition \citep{Kub01, Dew04, 
God09, Dew10}. The lower quality {\it ASCA} observations of two ULXs 
IC 342 X--1 and X--2 exhibited a canonical transition between low/hard and 
high/soft states \citep{Kub01}, where the two states were modeled 
with a PL and MCD component, respectively. However, there were flux 
variations not associated with any obvious state changes in some ULXs. 
For example, \swift{} monitoring observations of Holmberg II X--1, Holmberg IX X--1, 
NGC 5408 X--1 and NGC 4395 X--2 suggested that these ULXs remained in 
the same spectral states as their flux varied by an order of magnitude 
\citep{Kaa09, Gri10}. Variability studies have also revealed several 
correlations between the spectral parameters, when fitting relatively low 
counting statistics spectra with a MCD plus PL model, especially the disc 
luminosity versus temperature and photon index versus X-ray luminosity 
($\Gamma - L_{\rm X}$) correlations. Such correlations help compare 
the observed variations with the expectation from theoretical models. 
The disc luminosity versus temperature correlation, $L_{\rm Disc} 
\propto T^{4}$, has been reported for several ULXs \citep{Fen06, Fen09, Kaj09}, 
when modeled with absorbed MCD, which show consistency with the 
prediction from the standard accretion disc model. However, other 
ULXs appear to follow an anti-correlation between the parameters,
$L_{\rm Disc} \propto T_{\rm in}^{-3.5}$, when fitted with a cool MCD 
plus PL model \citep{Fen07, Sor07, Kaj09}. In addition, ULXs 
exhibited a $\Gamma-L_{\rm X}$ correlation when modeled with PL or MCD 
plus PL \citep[][and reference therein]{Fen06, Fen06a, Kaj08, Fen09, 
Kaj09}. Such a correlation has already been observed in BHXBs and active 
galactic nuclei \citep{Mag98, Zdz02, Zdz03}, and suggests that the 
variability mechanism is related to the geometry of the disc-corona in 
these sources \citep{Haa97, Mer01}.

The spectral variability studies of ULXs will thus help understand 
the geometry and physical processes in these systems, which can 
provide constraints on the nature of ULXs. A large number of 
available observations provide a unique opportunity for such studies 
on the nearby ULXs. In this paper, we present the long-term spectral 
variability study of the ULX Holmberg IX X--1 (hereafter Ho IX X--1) 
using the archival \suzaku{} and \xmm{} observations.
The data from the observations cover nearly the same energy range, 
0.3--10 keV, and are in good quality, allowing us to test different 
spectral models thoroughly and identify the model for well describing 
the source emission. In our study, we used a two-component 
thermal model (disc blackbody plus thermal Comptonization), 
which has been found to be able to describe the spectral variability of ULXs
quite well \citep{Vie10, Pin12, Pin14, Lua16}. 
In section \ref{hol}, we briefly explain Ho IX X--1 and its previous 
variability studies. The observations and data reduction method are 
described in the \S 2. The analysis and results are presented in \S 3. 
The discussion is presented in \S 4.

\subsection{Holmberg IX X--1 and the Variability Studies}
\label{hol}

Ho IX X--1 is one of the brightest ULXs with an average X-ray luminosity 
of $\sim 10^{40} \ergsec$. It is located near the galaxy M81 and is about 
2 arcmin away from the irregular dwarf galaxy Holmberg IX. Since the discovery 
with {\it Einstein} observatory \citep{Fab89}, 
the source has been studied with all major X-ray observatories \citep{Lap01}. 
The source was variable on time-scales of weeks to months and exhibited 
flux variations by a factor of seven in the {\it Swift} monitoring 
observations \citep{Kaa09}. Apart from the X-ray flux variability, 
Ho IX X--1 is one of the few ULXs that exhibited mHz QPOs in the long 
\xmm{} observation \citep{Dew06b}. The \xmm{} 
spectra were modeled by the cool accretion disc ($kT_{\rm in} \sim 
0.17 - 0.29$ keV) plus PL components and suggest an IMBH accretor 
of mass $\sim 10^3\,\rm M_\odot$ by scaling with the measured disc temperatures 
\citep{Mil04}. However, the {\it Swift} long-term monitoring observations 
of Ho IX X--1 \citep{Kon10} showed that the spectra can be described by 
the dual thermal model with cool blackbody and warm disc blackbody 
components. This model description leads to the suggestion that the 
accretor may be a $10\,\rm M_\odot$ black hole accreting well above the 
Eddington limit or a $100\,\rm M_\odot$ black hole at the Eddington rate. 
The broadband X-ray spectral study of Ho IX X--1, based on \suzaku{} 
and \xmm{} observations, confirmed the earlier indication of 
the cut-off feature in the spectra of bright ULXs \citep{Sto06}, which 
revealed the lack of an additional high-energy PL component that would 
otherwise contribute significantly to the X-ray emission \citep{Dew13}. 
Later, the coordinated broadband X-ray observations performed with \nustar{}, 
\xmm{} and \suzaku{} confirmed that the curvature observed 
previously in the limited bandpass is a true spectral cutoff \citep{Wal14}. 

Because Ho IX X--1 is one of the brightest nearby ULXs, its variability has 
been studied in the past. Earlier studies either used hardness ratios or 
the best-fitted disc plus PL model to investigate the spectral variability 
\citep{Lap01, Wan02, Mil04, Win07, Kaa09}. \cite{Vie10} studied 
the variability of the source in detail with a disc plus thermal Comptonization 
({\tt comptt}) model. Using the {\it Swift} and limited number of \xmm{} 
observations, they suggested a cool, optically thick Comptonizing corona 
for the source, which is consistent with the results reported in \cite{Gla09}. 
The observed variability is roughly characterized by a decrease in coronal 
electron temperature and an increase in the optical depth as the source 
becomes brighter. 

The coordinated broadband X-ray observations of Ho IX X--1 with \nustar{}, 
\xmm{} and \suzaku{} \citep{Wal14} revealed substantial spectral variability 
between two epochs of spectra (2012 October -- November). The broadband 
spectra, in the 0.3--30 keV energy band, were well described either by two 
thermal models ({\tt diskbb} and {\tt diskpbb}) or disc-corona model 
({\tt diskbb}+{\tt comptt}) along with a PL-like tail ({\tt simpl}) at high 
energies. The source was observed at a brighter state in the second epoch, where the 
flux and temperature of one of the thermal component changed significantly 
compared to the first epoch in the two thermal model description. 
The spectral evolution of the source could be dominated 
either by the hot or cool temperature components, while both scenarios 
require highly non-standard behavior in the observed evolution. 
Recently, four coordinated broadband observations with \suzaku{} and \nustar{} 
expanded the broadband variability study \citep{Wal17} and showed similar 
spectral variability. \cite{Lua16} studied the spectral evolution of Ho 
IX X--1 using the {\it Swift}, \nustar{}, and \xmm{} observations. 
They found that the flat or two-component spectra in the 1--6 keV band at 
lower luminosities tend to evolve to a curved and disc-like spectrum at higher 
luminosities and the peak energy in the curved spectrum decreases with increase 
in luminosity. They suggest that a super-critical accretion disc with 
massive winds can explain this spectral evolution of the source.

\section{Observations and Data Reduction}

We used the observations of Ho IX X--1 obtained with \suzaku{} 
and \xmm{}, which are publicly available, and analyzed the individual 
data in the 0.3--10 keV energy band. The list of the observations 
is given in Table~\ref{sample}.

\subsection{{\it Suzaku}}

Eight on-axis observations of Ho IX X--1 conducted with the \suzaku{} 
observatory \citep{Mit07} were used for the analysis. We reduced the 
unfiltered \suzaku{} data using the standard software package, 
{\sc heasoft} version 6.15.1, and reprocessed the X-ray imaging 
spectrometer (XIS) data using the specific {\sc headas} tool 
{\sc aepipeline}. The target was extracted from a circle of radius 
220 arcsec in XIS0, XIS1 and XIS3, whereas the background events 
were selected from two circular regions near the target region 
with radius of 110 arcsec. The front illuminated (FI) CCDs spectra, XIS0 
and XIS3, were added using the {\sc ftool addascaspec}. The co-added 
spectra were then grouped with a minimum of 60--300 counts per bin, 
depending upon the quality of the data.


\begin{table}
\tablecolumns{5}
\setlength{\tabcolsep}{10.0pt}
\tablewidth{320pt}
	\caption{Observation log in the chronological order}
 	\begin{tabular}{@{}lccc@{}}
	\hline
	\hline
\colhead{Data} & \colhead{ObsID} & \colhead{Date} & \colhead{Exposure} \\
     	       &		 & 		  & (ksec) \\
\hline

{\it XMM}1 & 0111800101 & 2001 Apr 22 & 132.7 \\
{\it XMM}2 & 0111800301 & 2001 Apr 22 & 8.0 \\
{\it XMM}3 & 0112521001 & 2002 Apr 10 & 10.7 \\
{\it XMM}4 & 0112521101 & 2002 Apr 16 & 11.5 \\
{\it XMM}5 & 0200980101 & 2004 Sep 26 & 119.1 \\
{\it XMM}6 & 0657802001 & 2011 Mar 24 & 27.5 \\
{\it XMM}7 & 0657801601 & 2011 Apr 17 & 21.1 \\
{\it XMM}8 & 0657801801 & 2011 Sep 26 & 25.4 \\ 
{\it XMM}9 & 0657802201 & 2011 Nov 23 & 23.9 \\
{\it Suzaku}1 & 707019010 & 2012 Apr 13 & 182.5 \\
{\it Suzaku}2 & 707019020 & 2012 Oct 21 & 107.5 \\
{\it XMM}10 & 0693850801 & 2012 Oct 23 & 14.1 \\
{\it Suzaku}3 & 707019030 & 2012 Oct 24 & 106.9 \\
{\it XMM}11 & 0693850901 & 2012 Oct 25 & 14.0 \\
{\it Suzaku}4 & 707019040 & 2012 Oct 26 & 110.0 \\
{\it XMM}12 & 0693851001 & 2012 Oct 27 & 13.9 \\
{\it XMM}13 & 0693851701 & 2012 Nov 12 & 9.9 \\
{\it XMM}14 & 0693851801 & 2012 Nov 14 & 13.8 \\
{\it XMM}15 & 0693851101 & 2012 Nov 16 & 13.3 \\
{\it Suzaku}5 & 709015010 & 2014 May 03 & 31.9 \\
{\it Suzaku}6 & 709015020 & 2014 Nov 15	& 34.1 \\
{\it Suzaku}7 & 709015030 & 2015 Apr 06 & 31.5 \\
{\it Suzaku}8 & 709015040 & 2015 May 16 & 34.1 \\

\hline
\end{tabular} 
\label{sample}
\end{table}

 
\subsection{{\it XMM-Newton}}

The \xmm{} data were obtained from the \xmm{} science archive and reduced 
using the standard tools of \xmm{} Science Analysis Software ({\sc sas}) 
version 14.0. The data from the \xmm{} European Photon Imaging Camera (EPIC) 
PN \citep{Str01} and metal oxide semiconductor \citep[MOS;][]{Tur01} detectors 
were used for the analysis. The full-field background light curve extracted 
from the EPIC camera in 10--12 keV energy range was used to select and 
remove the particle flaring background, and the good time intervals file 
was created. Out of fifteen \xmm{} observations, two observations, {\it XMM}\,2 
and {\it XMM}\,7 in Table~\ref{sample}, were highly affected by particle flaring 
and the available exposure time is too low to obtain good quality spectra. 
Thus, these observations were not included in the further analysis. We used 
the PN events with the best quality (FLAG = 0), PATTERN $ \le 4$, and removed 
the hot pixels in the data by using the flag expression $\#XMMEA\_EP$, while 
FLAG = 0, PATTERN $ \le 12$ and $\#XMMEA\_EM$ expression were used for the MOS 
data. The source and background events were selected from a circular region 
of radius ranges of 35--45 arcsec; the different extraction radii were for the 
purpose of avoiding chip gaps. The background regions were extracted from 
a source-free region near the ULX and if possible, from the same CCD. 
In some of the observations, especially {\it XMM}\,13, 
{\it XMM}\,14, and {\it XMM}\,15, the source was bright enough 
for EPIC-MOS to be affected by mild pileup \citep[see][]{Wal14}. To reduce the 
pileup effects, we considered only single grade events in these observations. 
The source and background spectra in the 0.3--10 keV band, along with response 
and ancillary response files, were extracted from the clean filtered event list 
using the standard \xmm{} {\sc sas} tasks. All spectra were grouped to minimum 
counts of 30--300, depending upon the quality of the data. 

\begin{table*}
\tablecolumns{6}
\setlength{\tabcolsep}{22.0pt}
\tablewidth{320pt}
	\caption{The obtained $\chi^2/\rm d.o.f$ for different models}
 	\begin{tabular}{@{}rccccl@{}}
	\hline
	\hline
\colhead{Data} & \colhead{Model 1} & \colhead{Model 2} & \colhead{Model 3} & \colhead{Model 4} & \colhead{Model 5}\\
\hline

{\it XMM}1 & $651.6/323$ & $582.4/321$ & $399.5/320$ & $388.6/320$ & $393.2/320$ \\ 
{\it XMM}3 & $347.4/288$ & $277.7/286$ & $267.2/285$ & $269.1/285$ & $269.6/285$ \\ 
{\it XMM}4 & $373.4/323$ & $318.4/321$ & $316.4/320$ & $315.5/320$ & $316.2/320$ \\ 
{\it XMM}5 & $1648.5/463$ & $541.6/461$ & $467.2/460$ & $463.4/460$ & $463.4/460$ \\ 
{\it XMM}6 & $279.0/230$ & $224.9/228$ & $224.3/227$ & $224.3/227$ & $224.4/227$ \\
{\it XMM}8 & $384.5/321$ & $377.9/319$ & $376.8/318$ & $373.6/318$ & $373.8/318$ \\ 
{\it XMM}9 & $438.7/392$ & $433.7/390$ & $397.7/389$ & $398.8/389$ & $399.8/389$ \\
{\it Suzaku}1 & $417.4/346$ & $403.2/344$ & $359.2/343$ & $356.7/343$ & $357.1/343$ \\
{\it Suzaku}2 & $346.5/289$ & $331.6/287$ & $313.3/286$ & $311.0/286$ & $312.8/286$ \\
{\it XMM}10 & $388.1/303$ & $317.3/301$ & $316.5/300$ & $315.7/300$ & $316.0/300$ \\
{\it Suzaku}3 & $315.0/300$ & $303.0/298$ & $300.1/297$ & $300.1/297$ & $301.0/297$ \\
{\it XMM}11 & $388.6/331$ & $319.0/329$ & $314.5/328$ & $313.1/328$ & $313.8/328$ \\
{\it Suzaku}4 & $345.8/359$ & $332.3/357$ & $326.7/356$ & $326.9/356$ & $327.1/356$ \\
{\it XMM}12 & $354.1/301$ & $334.3/299$ & $320.5/298$ & $319.3/298$ & $318.8/298$ \\
{\it XMM}13 & $498.6/366$ & $440.5/364$ & $363.8/363$ & $377.6/363$ & $378.4/363$ \\
{\it XMM}14 & $572.9/381$ & $511.2/379$ & $380.2/378$ & $379.7/378$ & $380.1/378$ \\
{\it XMM}15 & $454.9/333$ & $421.2/331$ & $370.2/330$ & $374.7/330$ & $374.5/330$ \\
{\it Suzaku}5 & $294.4/271$ & $293.2/269$ & $288.3/268$ & $288.2/268$ & $287.2/268$ \\
{\it Suzaku}6 & $261.6/253$ & $252.9/251$ & $233.7/250$ & $224.3/250$ & $224.4/250$ \\
{\it Suzaku}7 & $295.4/248$ & $279.9/246$ & $275.0/245$ & $276.0/245$ & $275.9/245$ \\
{\it Suzaku}8 & $267.6/246$ & $266.3/244$ & $266.3/243$ & $265.6/243$ & $266.0/243$ \\

\hline
\end{tabular} 
\tablecomments {Model 1: $\tt tbabs\times tbabs\times powerlaw$; Model 2: $\tt tbabs\times tbabs\times (diskbb + powerlaw)$; Model 3: $\tt tbabs\times tbabs\times (diskbb + cutoffpl)$; Model 4: $\tt tbabs\times tbabs\times (diskbb + compTT)$; Model 5: $\tt tbabs\times tbabs\times (diskbb + nthcomp)$.}
\label{chidof}
\end{table*}


\begin{table*}
\tablecolumns{10}
\setlength{\tabcolsep}{9.0pt}
\tablewidth{320pt}
	\caption{Best-fit parameters for the base-line model and the derived parameters from {\tt nthcomp} model function}
 	\begin{tabular}{@{}lccccccccc@{}}
	\hline
	\hline
\colhead{Data} & \colhead{$N_{\rm H}$} & \colhead{$kT_{\rm in}$} & \colhead{$\Gamma$} & \colhead{$kT_{\rm e}$} & \colhead{log $L_{\rm X}$} & \colhead{log $L_{\rm Disc}$} & \colhead{$\chi^2/\rm d.o.f$} & \colhead{log $L_{\rm Input}$} & \colhead{$A$} \\
\hline


{\it XMM}1 & $0.19^{+0.01}_{-0.01}$ & $0.20^{+0.01}_{-0.01}$ & $1.72^{+0.01}_{-0.01}$ & $>1.39$ & $40.41^{+0.01}_{-0.01}$ & $39.41^{+0.03}_{-0.03}$ & $393.2/320$ & $39.83$ & $3.90$ \\ 
{\it XMM}3 & $0.12^{+0.03}_{-0.03}$ & $0.27^{+0.05}_{-0.04}$ & $1.66^{+0.06}_{-0.08}$ & $2.38^{+0.74}_{-0.38}$ & $40.10^{+0.03}_{-0.03}$ & $39.36^{+0.10}_{-0.11}$ & $269.6/285$ & $39.44$ & $4.65$ \\ 
{\it XMM}4 & $0.17^{+0.04}_{-0.04}$ & $0.21^{+0.05}_{-0.03}$ & $1.81^{+0.04}_{-0.06}$ & $>2.56$ & $40.21^{+0.05}_{-0.04}$ & $39.39^{+0.21}_{-0.18}$ & $316.2/320$ & $39.61$ & $4.46$ \\ 
{\it XMM}5 & $0.13^{+0.01}_{-0.01}$ & $0.24^{+0.01}_{-0.01}$ & $1.55^{+0.02}_{-0.02}$ & $2.45^{+0.16}_{-0.13}$ & $40.05^{+0.01}_{-0.01}$ & $39.38^{+0.04}_{-0.04}$ & $463.4/460$ & $39.26$ & $6.25$ \\ 
{\it XMM}6 & $0.15^{+0.06}_{-0.05}$ & $0.25^{+0.06}_{-0.04}$ & $1.61^{+0.06}_{-0.09}$ & $>2.18$ & $40.17^{+0.06}_{-0.05}$ & $39.52^{+0.18}_{-0.15}$ & $224.4/227$ & $39.43$ & $6.05$ \\ 
{\it XMM}8 & $0.13^{+0.05}_{-0.03}$ & $0.23^{+0.11}_{-0.08}$ & $1.83^{+0.05}_{-0.04}$ & $>3.01$ & $40.30^{+0.05}_{-0.03}$ & $<39.20$ & $373.8/318$ & $39.79$ & $3.97$ \\
{\it XMM}9 & $0.14^{+0.04}_{-0.03}$ & $0.25^{+0.08}_{-0.06}$ & $1.80^{+0.03}_{-0.04}$ & $2.22^{+0.32}_{-0.23}$ & $40.42^{+0.03}_{-0.02}$ & $39.02^{+0.31}_{-1.67}$ & $399.8/389$ & $39.91$ & $3.68$ \\ 
{\it Suzaku}1 & $<0.09$ & $0.30^{+0.10}_{-0.08}$ & $1.64^{+0.03}_{-0.05}$ & $2.57^{+0.25}_{-0.12}$ & $40.16^{+0.04}_{-0.02}$ & $38.91^{+0.28}_{-0.29}$ & $357.1/343$ & $39.55$ & $4.76$ \\ 
{\it Suzaku}2 & $<0.19$ & $0.21^{+0.12}_{-0.04}$ & $1.70^{+0.04}_{-0.04}$ & $2.76^{+0.45}_{-0.29}$ & $40.23^{+0.09}_{-0.06}$ & $39.18^{+0.47}_{-0.87}$ & $312.7/286$ & $39.59$ & $5.05$ \\  
{\it XMM}10 & $0.14^{+0.04}_{-0.03}$ & $0.24^{+0.05}_{-0.04}$ & $1.68^{+0.04}_{-0.06}$ & $>2.56$ & $40.21^{+0.04}_{-0.03}$ & $39.40^{+0.16}_{-0.14}$ & $316.0/300$ & $39.54$ & $5.43$ \\  
{\it Suzaku}3 & $<0.17$ & $0.23^{+0.16}_{-0.11}$ & $1.75^{+0.04}_{-0.04}$ & $3.03^{+0.75}_{-0.37}$ & $40.23^{+0.10}_{-0.03}$ & $<39.55$ & $301.0/297$ & $39.67$ & $4.58$ \\  
{\it XMM}11 & $0.13^{+0.03}_{-0.03}$ & $0.28^{+0.06}_{-0.05}$ & $1.68^{+0.05}_{-0.07}$ & $2.94^{+1.78}_{-0.59}$ & $40.24^{+0.03}_{-0.02}$ & $39.43^{+0.10}_{-0.11}$ & $313.8/328$ & $39.60$ &$4.80$ \\ 
{\it Suzaku}4 & $0.17^{+0.09}_{-0.12}$ & $0.17^{+0.06}_{-0.05}$ & $1.80^{+0.04}_{-0.05}$ & $3.45^{+1.53}_{-0.59}$ & $40.33^{+0.11}_{-0.10}$ & $39.53^{+0.39}_{-0.90}$ & $327.1/356$ & $39.70$ & $4.81$ \\ 
{\it XMM}12 & $0.10^{+0.04}_{-0.03}$ & $0.31^{+0.10}_{-0.08}$ & $1.67^{+0.09}_{-0.15}$ & $2.16^{+0.53}_{-0.36}$ & $40.24^{+0.04}_{-0.03}$ & $39.39^{+0.17}_{-0.18}$ & $318.8/298$ & $39.63$ & $4.13$ \\ 
{\it XMM}13 & $0.22^{+0.09}_{-0.13}$ & $0.16^{+0.28}_{-0.03}$ & $1.73^{+0.06}_{-0.05}$ & $1.88^{+0.18}_{-0.13}$ & $40.52^{+0.12}_{-0.10}$ & $<40.08$ & $378.4/363$ & $39.89$ & $4.59$ \\ 
{\it XMM}14 & $0.17^{+0.06}_{-0.05}$ & $0.20^{+0.07}_{-0.03}$ & $1.62^{+0.04}_{-0.03}$ & $1.66^{+0.11}_{-0.09}$ & $40.51^{+0.06}_{-0.04}$ & $39.43^{+0.36}_{-0.51}$ & $380.0/378$ & $39.84$ & $4.91$ \\ 
{\it XMM}15 & $0.22^{+0.09}_{-0.09}$ & $0.17^{+0.07}_{-0.02}$ & $1.67^{+0.06}_{-0.06}$ & $1.72^{+0.20}_{-0.14}$ & $40.56^{+0.11}_{-0.08}$ & $39.68^{+0.46}_{-1.18}$ & $374.5/330$ & $39.87$ & $4.86$ \\
{\it Suzaku}5 & $<0.15$ & $0.39^{+0.12}_{-0.29}$ & $1.65^{+0.09}_{-0.17}$ & $2.63^{+1.01}_{-0.55}$ & $40.15^{+0.11}_{-0.01}$ & $<39.46$ & $287.2/268$ & $39.59$ & $4.20$ \\ 
{\it Suzaku}6 & $<0.13$ & $0.36^{+0.14}_{-0.10}$ & $1.44^{+0.10}_{-0.20}$ & $1.96^{+0.30}_{-0.29}$ & $40.05^{+0.05}_{-0.02}$ & $39.28^{+0.15}_{-0.19}$ & $224.4/250$ & $39.29$ & $5.69$ \\ 
{\it Suzaku}7 & $0.25^{+0.17}_{-0.19}$ & $0.19^{+0.08}_{-0.03}$ & $1.61^{+0.07}_{-0.08}$ & $2.62^{+1.18}_{-0.47}$ & $40.18^{+0.22}_{-0.15}$ & $39.70^{+0.45}_{-0.61}$ & $275.8/245$ & $39.32$ & $6.27$ \\ 
{\it Suzaku}8 & $<0.35$ & $<0.48$ & $1.81^{+0.06}_{-0.06}$ & $>2.41$ & $40.20^{+0.22}_{-0.14}$ & $<40.07$ & $266.0/243$ & $39.59$ & $5.05$ \\

\hline
\end{tabular} 
\tablecomments {(1) Data; (2) neutral hydrogen column density in units of $10^{22}~\rm cm^{-2}$; (3) inner disc temperature in keV; (4) photon index; (5) electron temperature in keV; (6) - (7) logarithmic total unabsorbed 0.3--10 keV X-ray luminosity and the luminosity of disc component in $\ergsec$, calculated by assuming the distance of 3.4\mpc~\citep{Hil93}; (8) $\chi^2$ statistics and degrees of freedom; (9) logarithmic luminosity of input seed photons in $\ergsec$; (10) amplification factor. Seed photon temperature, $kT_{\rm S} = kT_{\rm in}$ and input type = 1.}
\label{model7}
\end{table*}


\section{Analysis and Results}

\subsection{Spectral Analysis}

We performed detailed spectral modeling with simple models and complex 
physical models to understand the variability. The spectral modeling were 
performed with {\sc xspec} version 12.8.1g \citep{Arn96}. We fitted the 
\suzaku{} and \xmm{} (PN and MOS simultaneously) spectra in the 
0.6--10 keV and 0.3--10 keV, respectively. 
In order to perform the spectral analysis in the same energy 
band, i.e., 0.3--10 keV, we extended the low-energy range of \suzaku{} 
data to 0.3 keV (using {\sc energies} command in {\sc xspec}) 
and derived the spectral parameters based on the best-fit models. 
Due to the calibration uncertainties, the 1.7--2 keV 
energy range was excluded from the \suzaku{} XIS spectra. 
The uncertainties on the best-fit parameters were quoted at a 90\% 
confidence level. Two multiplicative absorption components \citep[{\tt tbabs} in {\sc xspec};][]
{Wil00} were used to incorporate the intervening absorption. The first 
component was fixed at the Galactic column $N_{\rm H,Gal}=5.54\times10^{20}~\rm cm^{-2}$ 
\citep{Kal05} towards the direction of the source, while the second component, 
which was free to vary, represents the absorption local to the ULX. 

Initially, we fitted the spectra with an absorbed PL model. 
The model provides a statistically acceptable fit 
for only six spectra, where the null hypothesis probability $> 0.05$,
and fails to explain the spectra from {\it XMM}\,1 and {\it XMM}\,5,
where the reduced $\chi^2 > 2$ ($\chi^2$/degrees of freedom (d.o.f) = 
651.6/323 and 1648.5/463, respectively; see Table~\ref{chidof}). Thus, 
we added the MCD model ({\tt diskbb} in {\sc xspec}) to the PL. This 
combined model improved the fit significantly in majority of 
the spectra compared to the absorbed PL and the differences in the 
$\chi^2$ values of the two models ranges from 1106.9 (obs. {\it XMM}\,5) to 1.2 (obs. {\it Suzaku}\,5) 
for the loss of two additional d.o.f (see Table~\ref{chidof}).
In addition, improved fits were obtained for {\it XMM}\,1 and {\it XMM}\,5 spectra 
($\chi^2$/d.o.f = 582.4/321 and 541.6/461, respectively). 
We then attempted with an exponentially cut-off PL ({\tt cutoffpl}) by 
replacing the PL component, which further improved the fit, especially for the \xmm{} 
observations, though the cut-off energies ($E_{\rm cutoff} > 3$ keV) were 
not well constrained in some of the cases. As pointed out in \S 1, 
the high energy cut-off or curvature commonly found in the bright ULXs can 
be described by thermal Comptonization in a cool and optically thick corona. 
Using thermal Comptonization models, {\tt comptt} or {\tt nthcomp} 
(in {\sc xspec}), is appropriate in such circumstances. Thus, we tested 
these models by replacing the {\tt cutoffpl} and both the models provide 
statistically acceptable fit to the spectra (see Table~\ref{chidof}). 
While the former model is widely used to characterize the thermal 
Comptonization, here we proceeded with the {\tt nthcomp} model to 
explain the observed spectral features. 

The {\tt nthcomp} model \citep{Zdz96, Zyc99} describes the thermal 
Comptonization in a relatively cool and optically thick plasma. The 
electron plasma temperature ($kT_{\rm e}$), photon index ($\Gamma$) and 
seed photon temperature ($kT_{\rm S}$) are the free parameters of 
the model. The soft X-ray emission described by MCD is considered 
as the principal component, providing seed photons for the 
Comptonization. Thus, the temperatures of the MCD component ($kT_{\rm in}$) and the 
seed photons were kept the same and varied together. In the disc-corona 
models, if the optically thick corona masks the underneath disc, the 
seed photon temperature is not always equal to the disc temperature. 
We also repeated the analysis by disconnecting the two temperatures. 
However this assumption led to difficulties in finding a unique global 
minimum in the spectral fit \citep[see also][]{Fen09, Pin12}. Thus 
we decided to keep the two temperatures as the same. Because the 
${\tt tbabs\times tbabs\times (diskbb + nthcomp)}$ model provides 
the acceptable best-fits and reasonable physical explanations, 
we considered it as the base-line model. All the spectra were 
fitted with this base-line model and the best-fit parameters are 
listed in Table~\ref{model7}. The total unabsorbed luminosity ($L_{\rm X}$) 
and the luminosity of disc component ($L_{\rm Disc}$) were calculated using 
the convolution model {\tt cflux} available in {\sc xspec}.

\subsection{Spectral Variability}

\begin{figure*}
\begin{center}

 \includegraphics[width=7.7cm,angle=0]{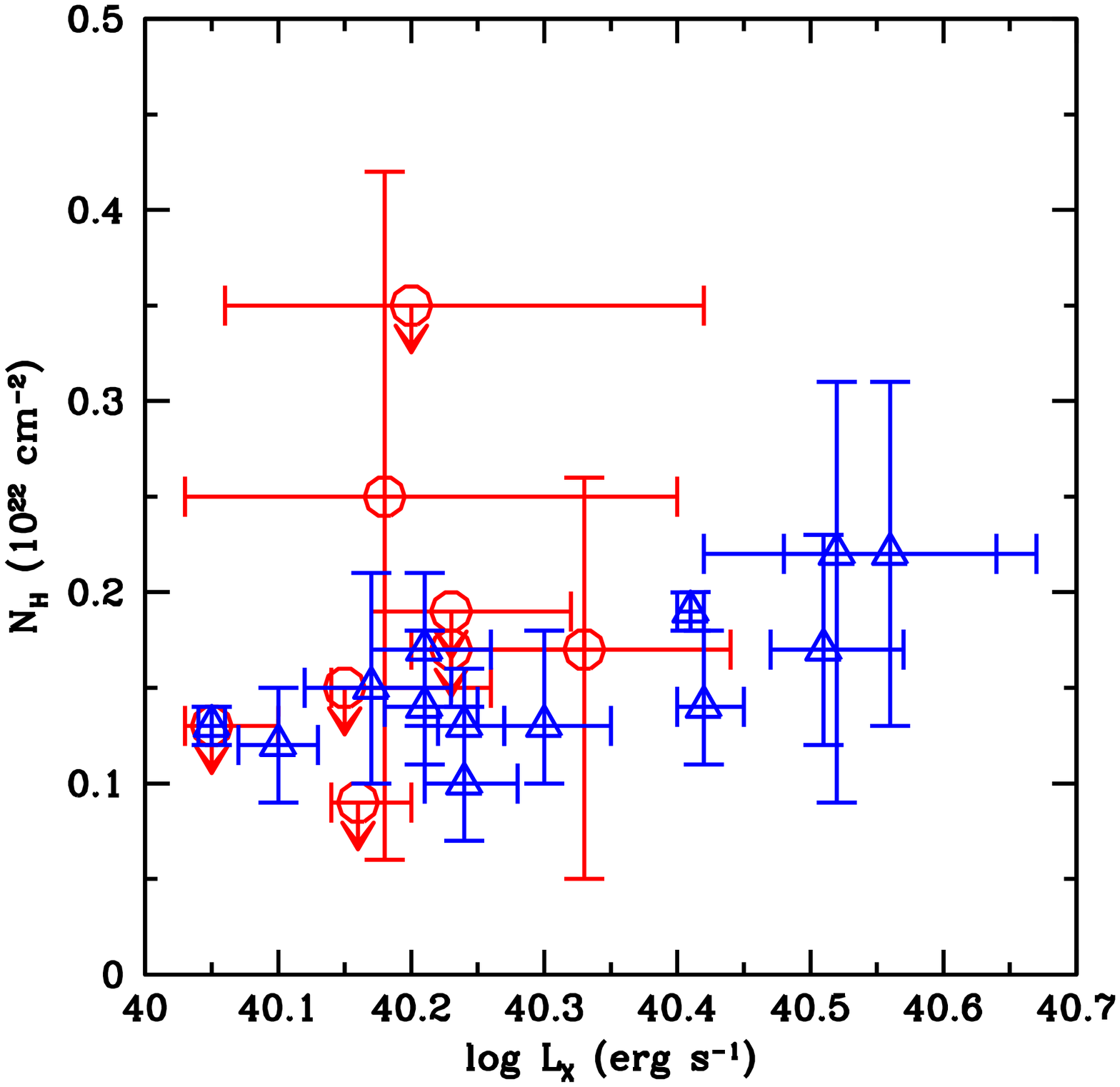}
 \includegraphics[width=7.7cm,angle=0]{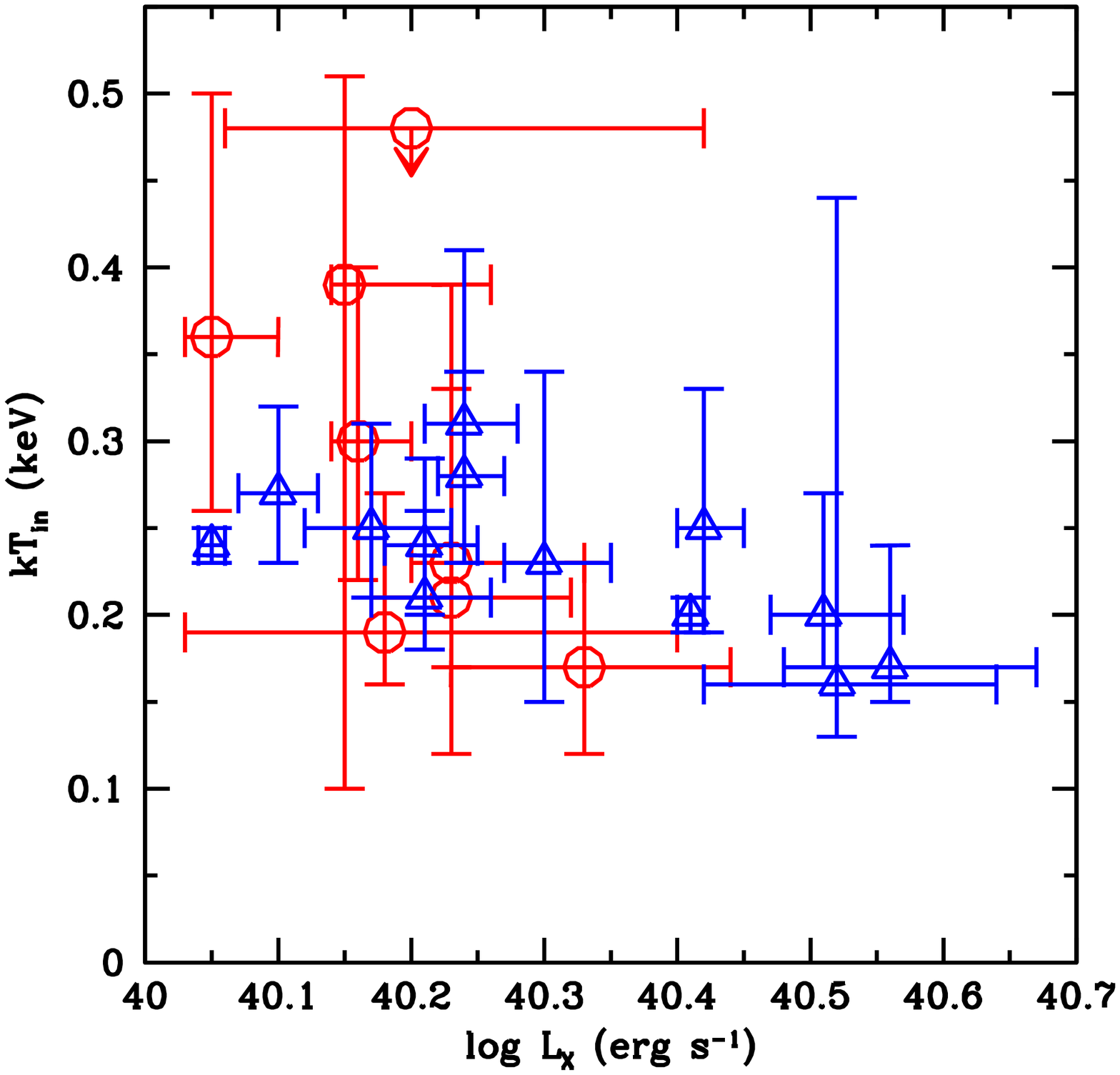}

 \includegraphics[width=7.7cm,angle=0]{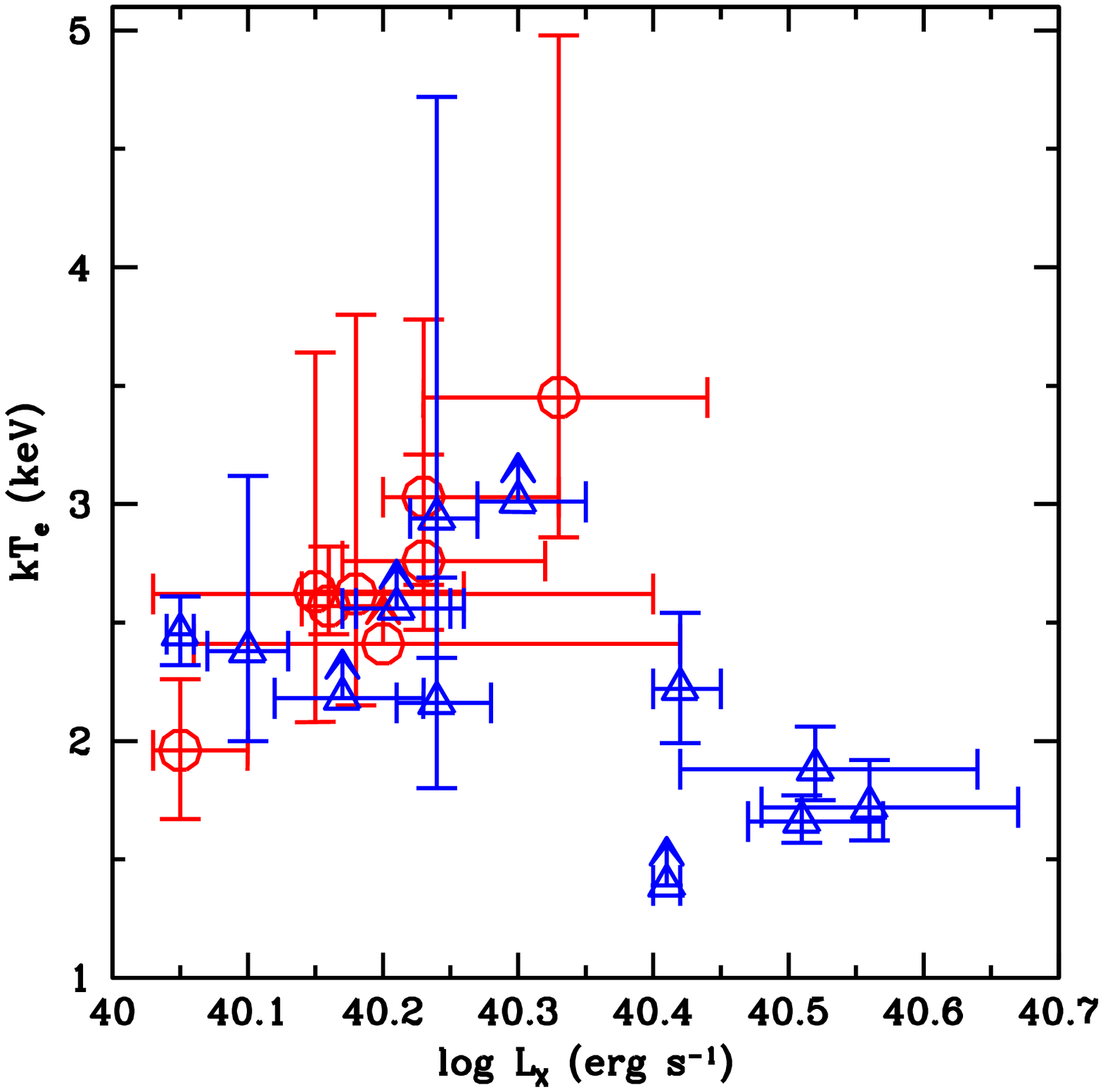}
 \includegraphics[width=7.7cm,angle=0]{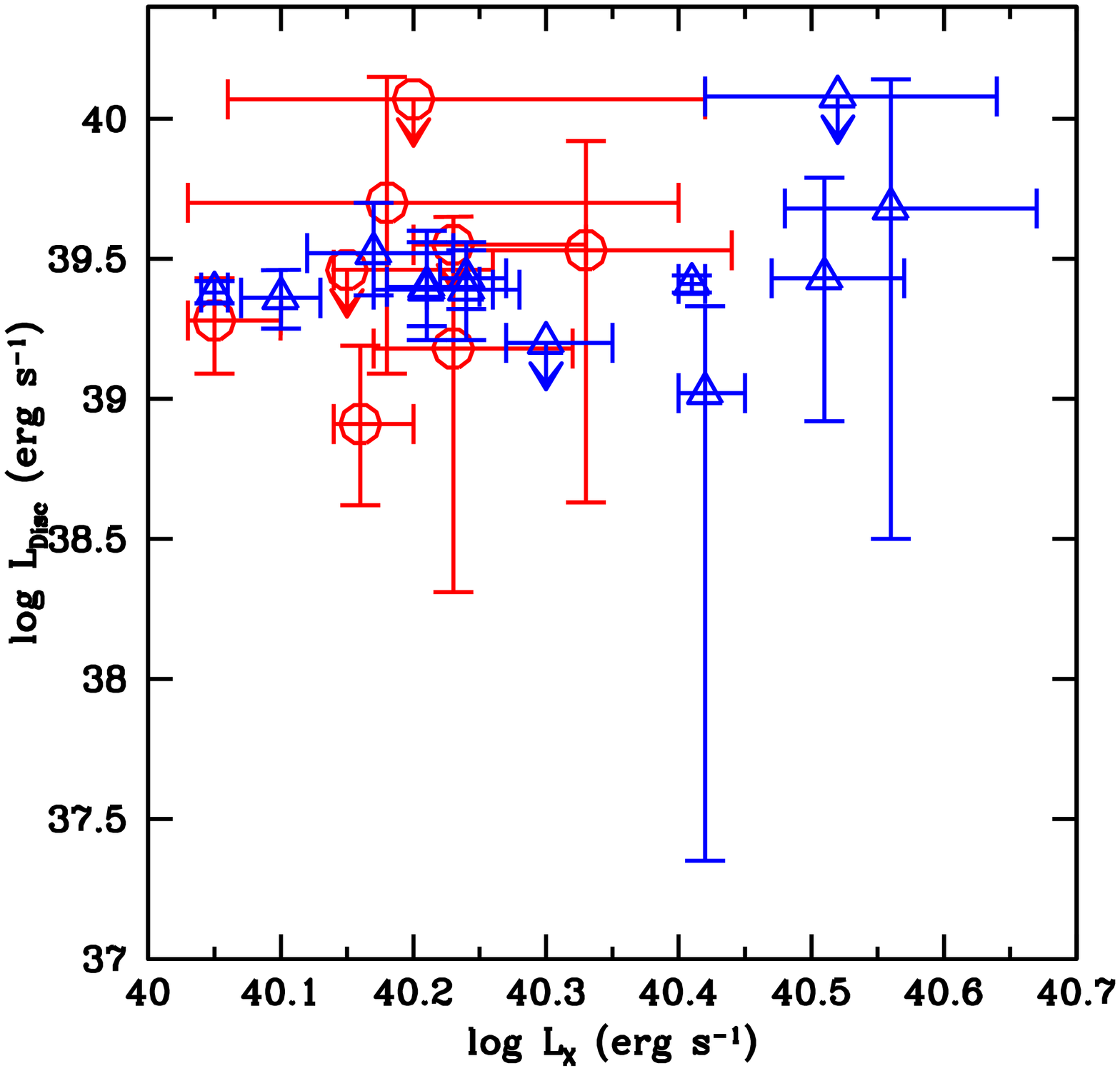}

 \includegraphics[width=7.7cm,angle=0]{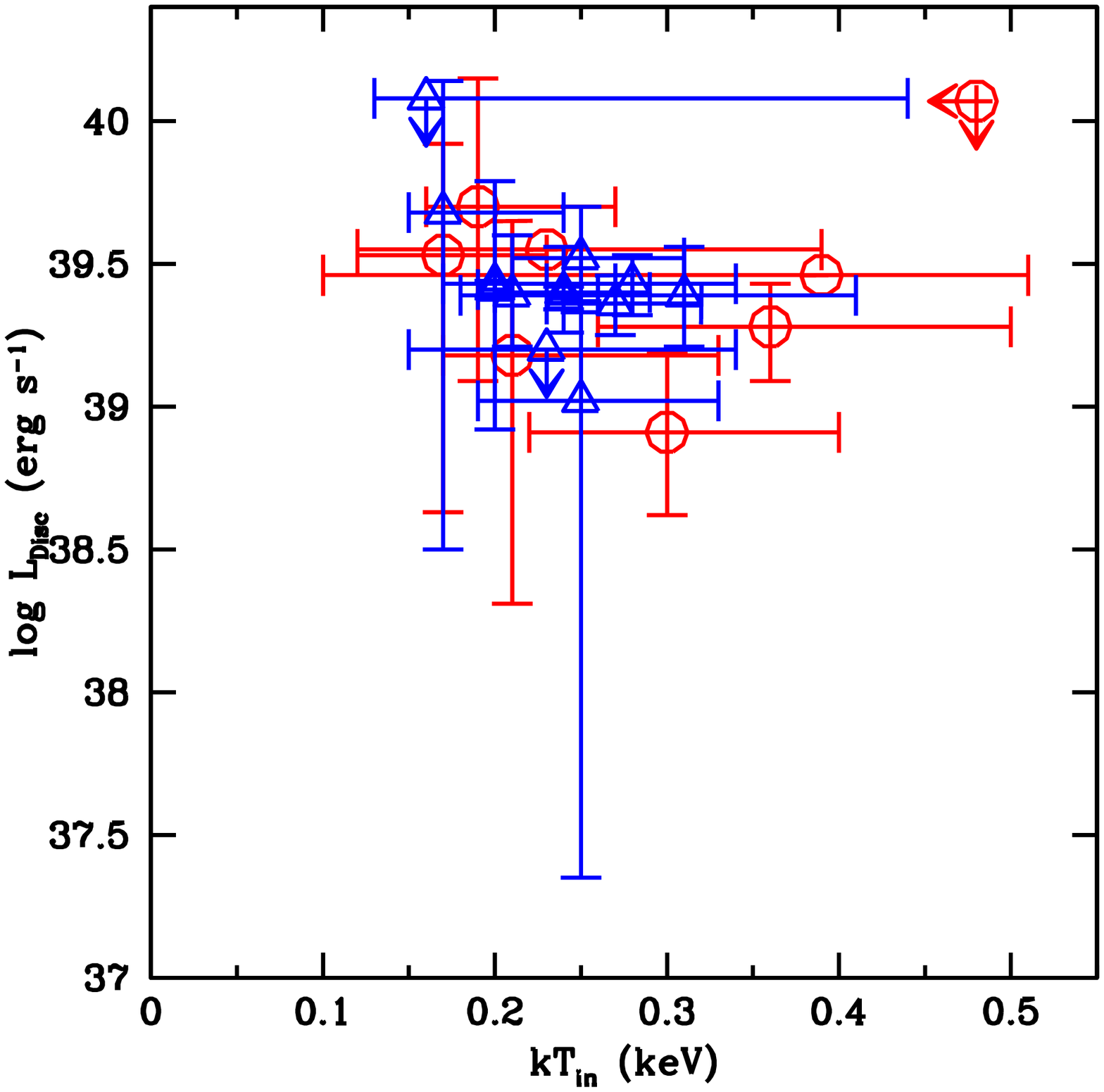}

\caption{The variations of best-fit spectral parameters ($N_{\rm H}$, $kT_{\rm in}$ and $kT_{\rm e}$) with X-ray luminosity ($L_{\rm X}$ and $L_{\rm Disc}$) in the 0.3--10~keV energy band. The red open circles and blue triangles represent the \suzaku{} and \xmm{} data respectively.}
\label{lxVsparams}
\end{center}
\end{figure*}

\begin{figure}
\begin{center}

\includegraphics[width=8.5cm,angle=0]{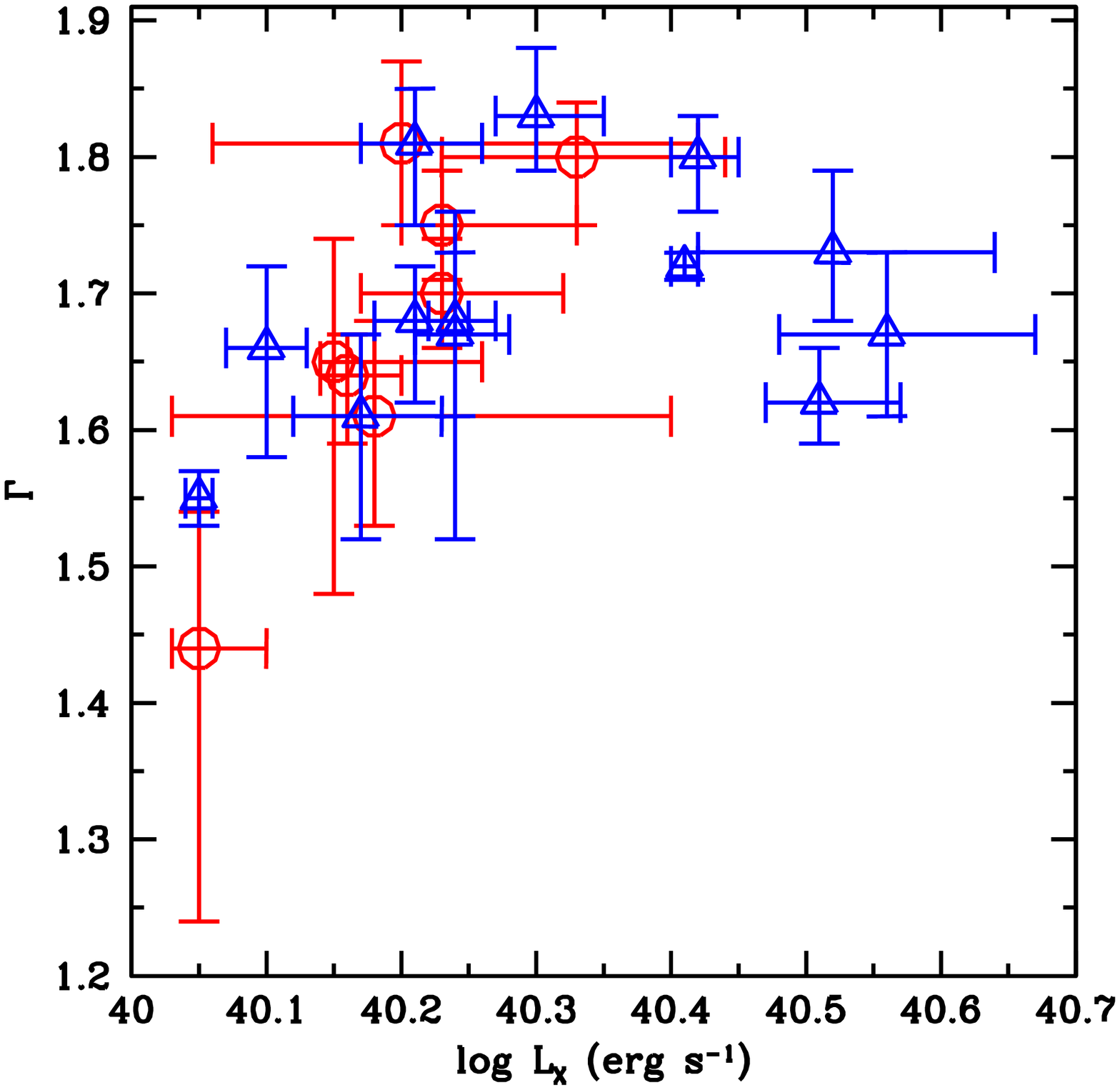}

\caption{Photon index versus X-ray luminosity for Ho IX X--1 obtained from the base-line model. The red open circles and blue triangles represent the \suzaku{} and \xmm{} spectra respectively.}
\label{lxVsgamma}
\end{center}
\end{figure}

We first investigated how the spectral parameters, $N_{\rm H}$, 
$kT_{\rm in}$, $kT_{\rm e}$, $L_{\rm Disc}$, and $\Gamma$, 
changed with X-ray luminosity $L_{\rm X}$. Their variations 
as a function of $L_{\rm X}$ are shown in Figure~\ref{lxVsparams} 
and \ref{lxVsgamma}. From the figures, it is clear that the 
parameters exhibit an approximate trend with $L_{\rm X}$. 
Among them, the plasma temperature ($kT_{\rm e} \sim 1-4$ keV) has large 
uncertainties and was poorly constrained in some of the observations, 
although it first increases, and then decrease as $L_{\rm X}$ increases. 
The disc luminosity $L_{\rm Disc}$ is in the range of a few times $10^{39} 
\ergsec$ with large uncertainties and not well constrained in 
some of the observations. Thus, we can consider $L_{\rm Disc}$ not
strongly variable or relatively stable in these observations. 
The photon index $\Gamma$ varies significantly with $L_{\rm X}$ 
(Figure~\ref{lxVsgamma}). It first has a strong correlation with 
$L_{\rm X}$, but turns to be lower at the highest $L_{\rm X}$. 

We quantified the correlations between the X-ray luminosity and 
the best-fit parameters by obtaining the Spearman rank correlation 
coefficient ($r_{\rm s}$) and the probability ($p$-value) for the null 
hypothesis. If the $p$-value is 0.05 or less, the correlation is 
considered to be significant. The parameters $N_{\rm H}$ and $\Gamma$ 
are positively correlated with $L_{\rm X}$, where $r_{\rm s}=0.53$ 
($p=1.43\times10^{-2}$) and $r_{\rm s}=0.52$ ($p=1.51\times10^{-2}$), respectively. 
The parameter $kT_{\rm in}$ is negatively correlated with both $L_{\rm X}$ and 
$L_{\rm Disc}$, where $r_{\rm s}=-0.58$ ($p=5.57\times10^{-3}$) and 
$r_{\rm s}=-0.61$ ($p=3.41\times10^{-3}$), respectively. However, the 
parameters $L_{\rm Disc}$ and $kT_{\rm e}$ do not show any significant 
correlation with $L_{\rm X}$ and the coefficients are $r_{\rm s}=0.31$ 
($p=0.17$) and $r_{\rm s}=-0.30$ ($p=0.18$), respectively. 

We note that $N_{\rm H}$ is positively correlated with $L_{\rm X}$. 
If the correlation has a physical origin, the increasing luminosity is a 
signature of the denser environment, while a non-physical origin 
is due to the degeneracy between $N_{\rm H}$ and $\Gamma$ in model fitting,
artificially boosting the unabsorbed luminosity 
\citep[see more details in][]{Kaj09}. For some of the ULXs, $N_{\rm H}$ 
is also correlated with $\Gamma$, when fitted with a PL or a disc plus 
PL model \citep {Fen09, Kaj09}. However, we did not find an $N_{\rm H}$--$\Gamma$ 
correlation for Ho IX X--1 with the base-line model. This result is 
consistent with that in \cite{Kaj09}, where the $N_{\rm H}-\Gamma$ correlation 
no longer exists when fitted with the PL plus cool MCD model. Since 
$N_{\rm H}$ varies marginally with $L_{\rm X}$, we tried to fix the 
absorption column to their mean value ($N_{\rm H}=1.34\times10^{21}~\rm cm^{-2}$) 
and fitted the spectra again. The spectral fits were worse for the 
cases where the best-fit values of $N_{\rm H}$ differ from the mean 
value, but we found that the $\Gamma-L_{\rm X}$ correlation still 
appears. Thus, although we considered the local absorption column 
as a free parameter in the base-line model, this choice 
did not affect the correlation (for a given spectrum, higher 
$N_{\rm H}$ can cause higher $\Gamma$ and higher $L_{\rm X}$).

Examining the correlation between $\Gamma$ and $L_{\rm X}$ 
(Figure~\ref{lxVsgamma}), $\Gamma$ reaches $\sim$1.8 
around the luminosity of $2\times10^{40}\,\ergsec$ and actually
turns to be lower with values of $\sim$1.6--1.7 at higher luminosities.
We investigated this plausible turn-over feature by fitting a line to 
all data points first.  A poor fit with a reduced $\chi^2$ of 3.6 
(19 d.o.f) was obtained.
Excluding the data points above the luminosity of $2\times10^{40}\,\ergsec$,
then the fitting was improved to have a reduced $\chi^2$ of 1.0 (13 d.o.f). 
This improvement is significant at a $99.95\%$ confidence level
compared to the first fit, indicating the presence of the turn-over.
We also attempted a broken power law fit for the $\Gamma$ 
and $L_{\rm X}$ correlation. The fitting identified a break at 
$1.99\pm0.10~\times10^{40}\,\ergsec$ with a reduced $\chi^2$ of $\sim 1.7$  
for 17 d.o.f, and the slopes were $0.27\pm0.04$ and $-0.17\pm0.05$ below 
and above the break respectively. The fit results further support a 
turn-over around $\sim 2\times10^{40}\,\ergsec$.

Many ULXs, when their spectra were fitted with a PL plus cool MCD model,
exhibited an anti-correlation between the disc luminosity and 
temperature, $L_{\rm Disc} \propto T_{\rm in}^{-3.5}$ \citep {Sor07, Kaj09}.
This anti-correlation suggests that the ULXs considered in those work 
were in a state of having high-accretion rate and low-disc temperature 
\citep[ultraluminous branch;][]{Sor07}. We checked the correlation 
between $L_{\rm Disc}$ and $kT_{\rm in}$ derived from the base-line 
model, and the anti-correlation appears to be consistent with the 
previous reported results \citep {Sor07, Kaj09}. 

The X-ray luminosity of Ho IX X--1 has changed by a factor of 
$\sim 3-4$ during these observations and among them, {\it XMM}\,5 
and {\it Suzaku}\,6 data have the lowest flux. During the observations 
conducted in 2012 November ({\it XMM}\,13 -- {\it XMM}\,15), the source 
was observed with the highest X-ray luminosity ($L_{\rm X} \sim 3.6\times 10^{40} \ergsec$) 
compared to the other observations. The MCD component appears relatively 
less variable and the flux contribution from it is always less ($\lesssim 
33$\%) than that from the Comptonized component in all observations. 
Thus, the observed variability is dominated by the Comptonization process.

\begin{figure*}
\begin{center}

\includegraphics[width=7.7cm,angle=0]{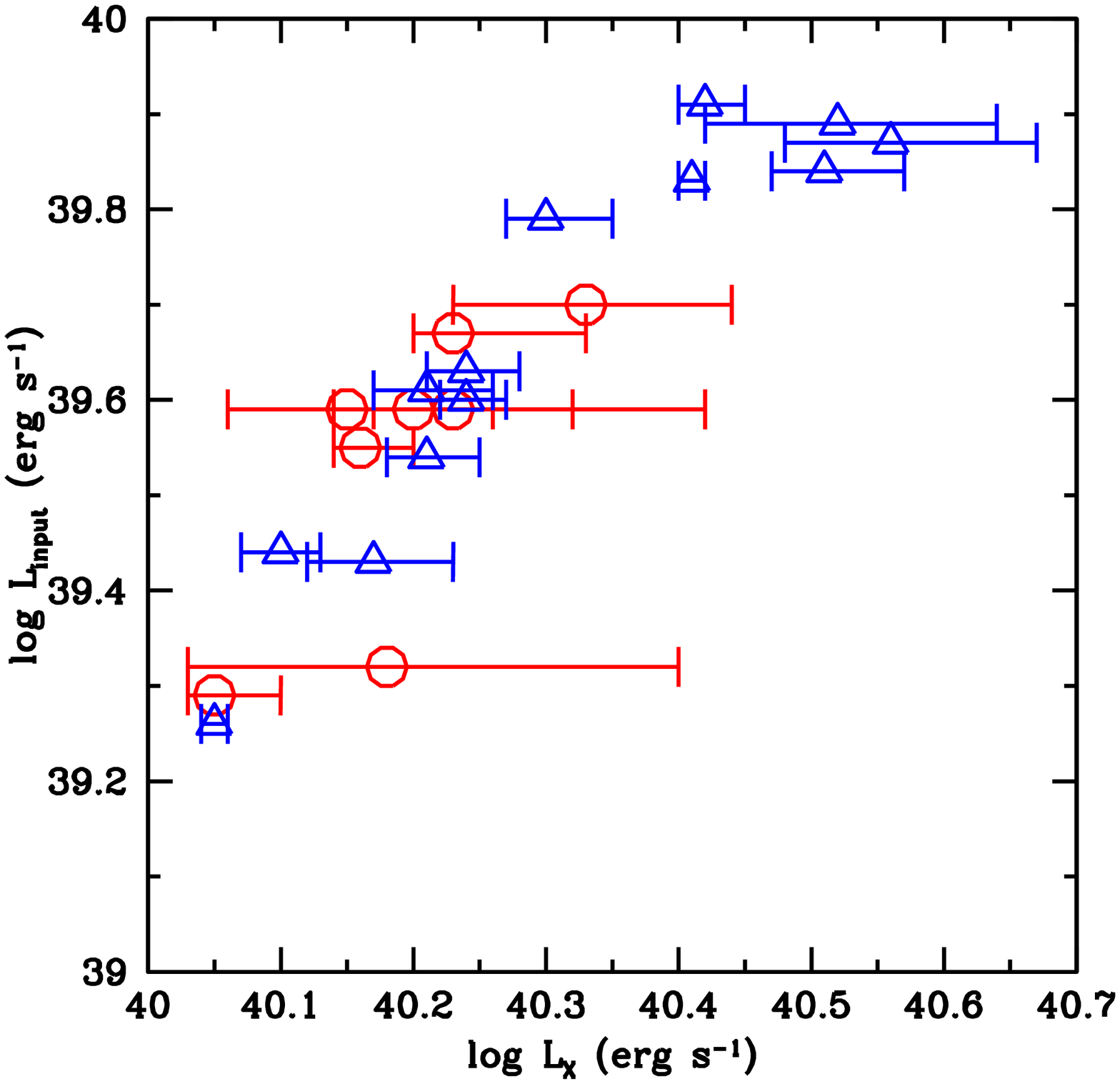}
\includegraphics[width=7.7cm,angle=0]{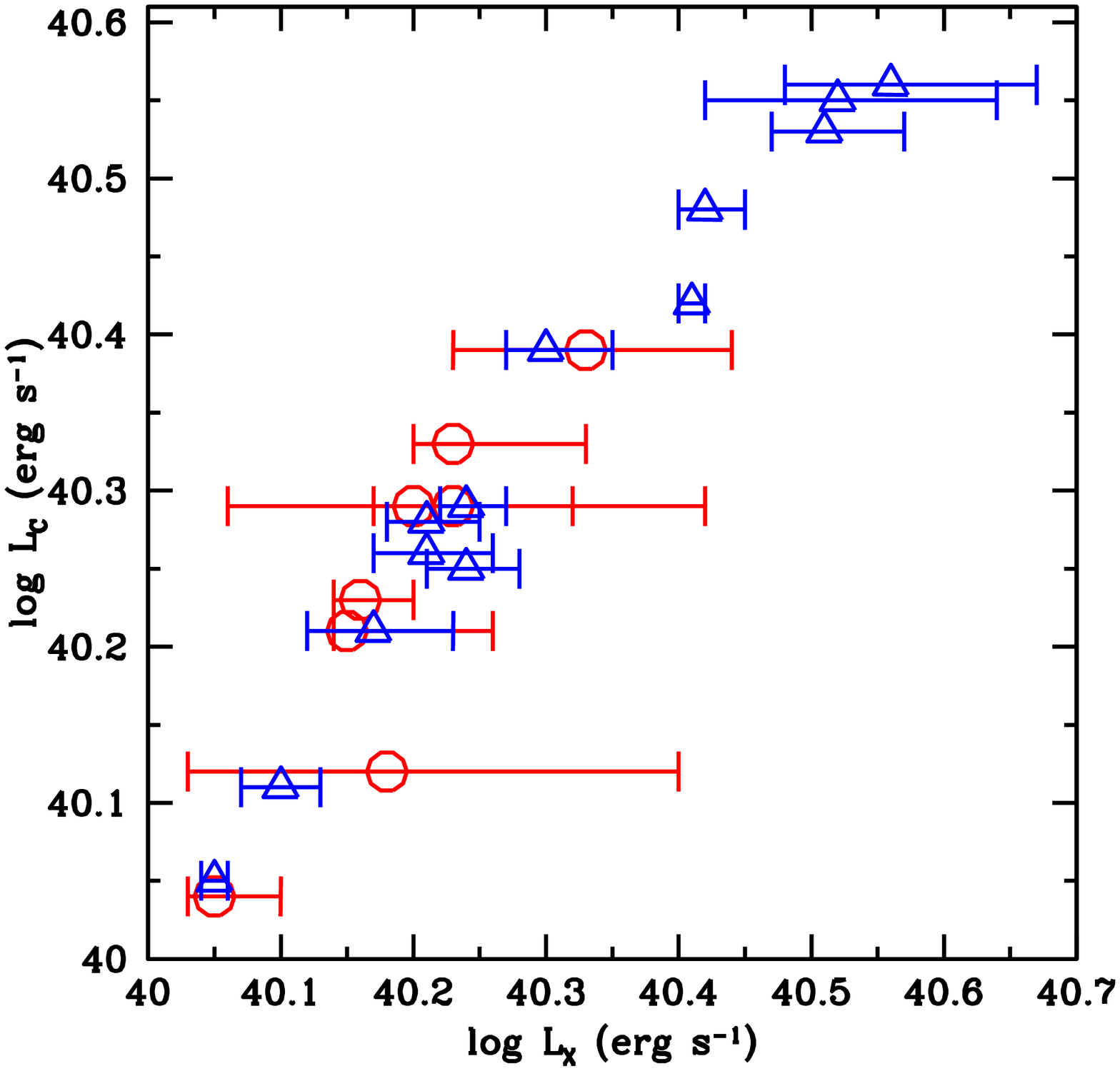}

\includegraphics[width=7.7cm,angle=0]{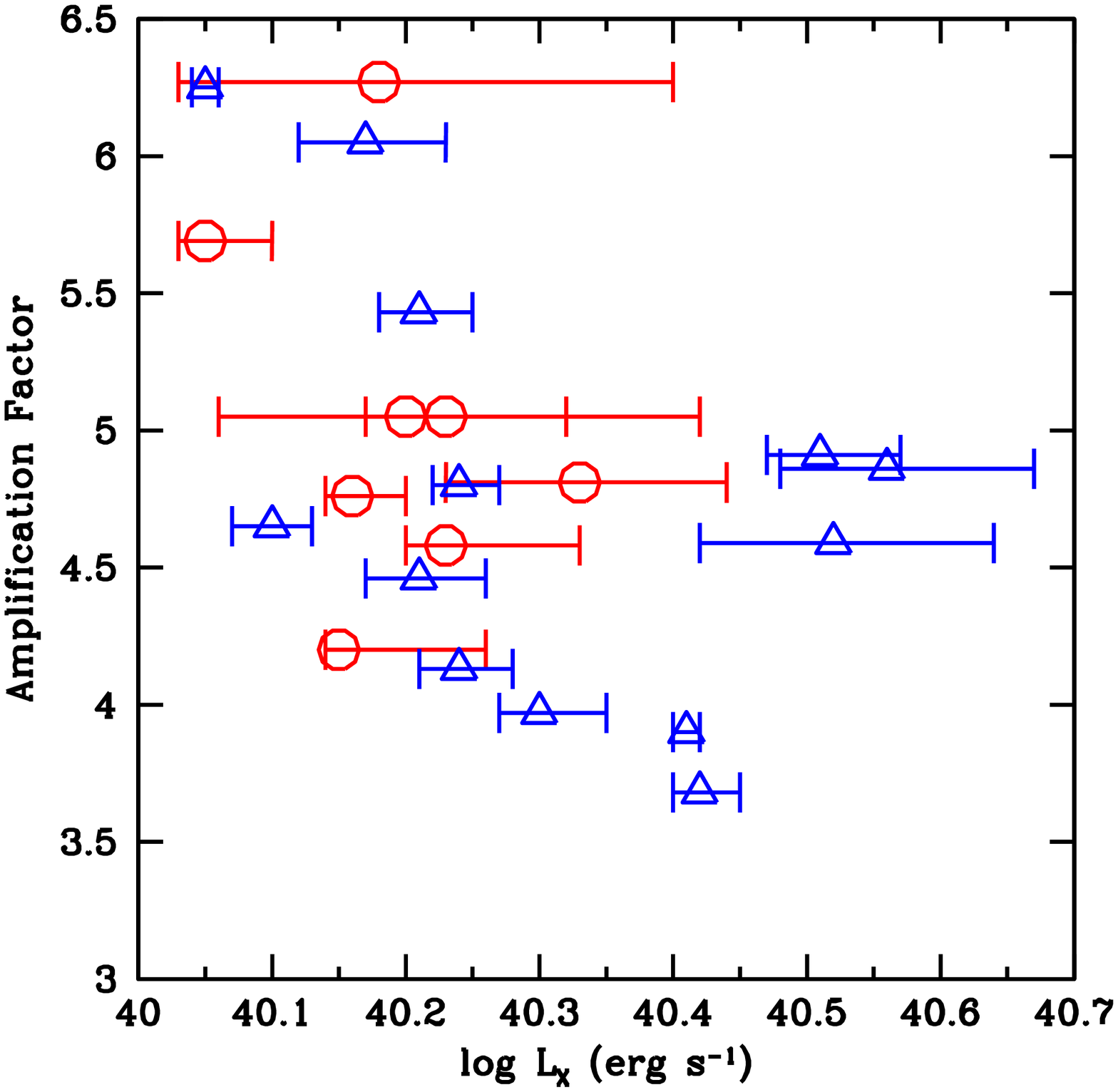}
\includegraphics[width=7.7cm,angle=0]{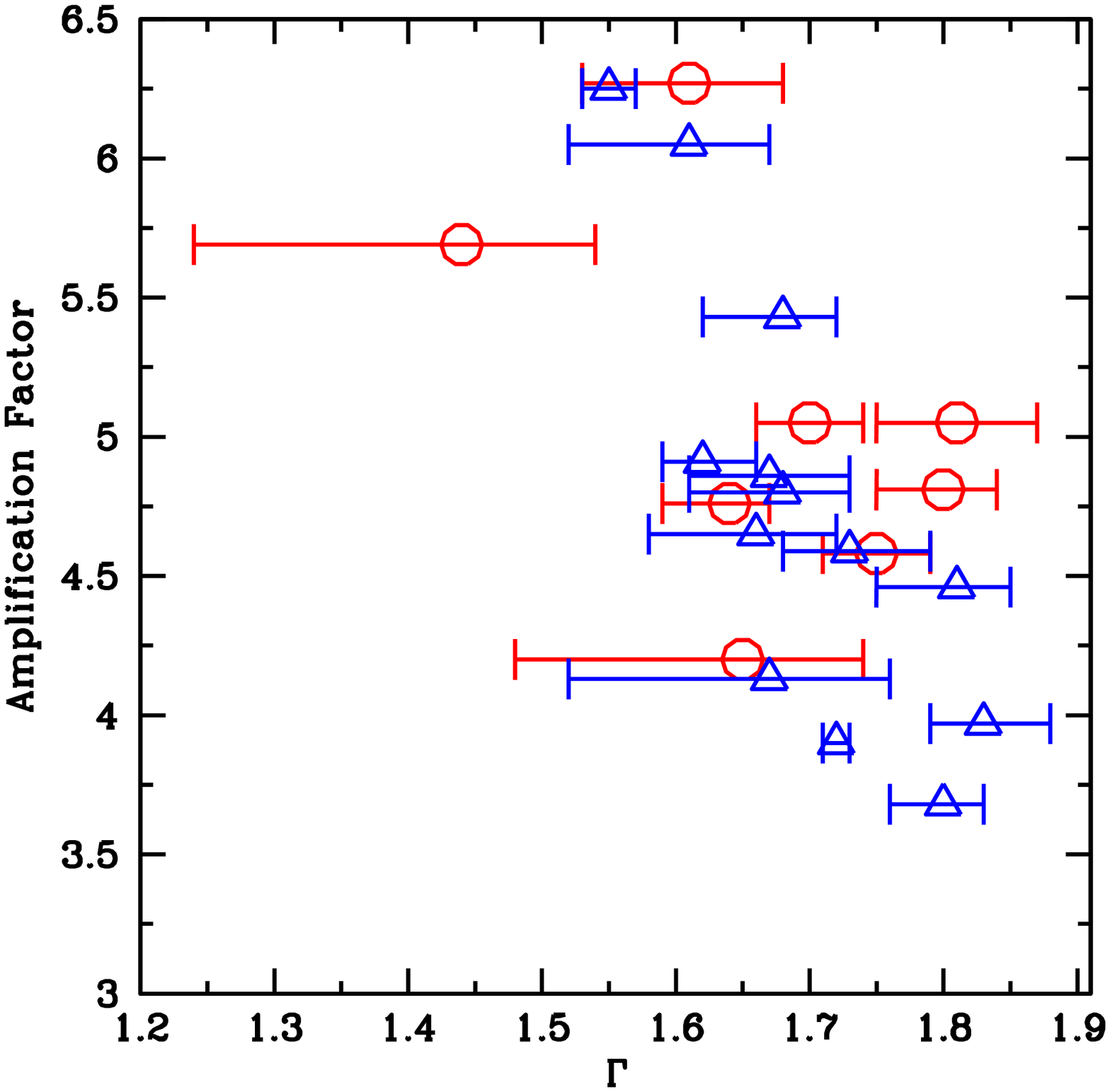}

\includegraphics[width=7.7cm,angle=0]{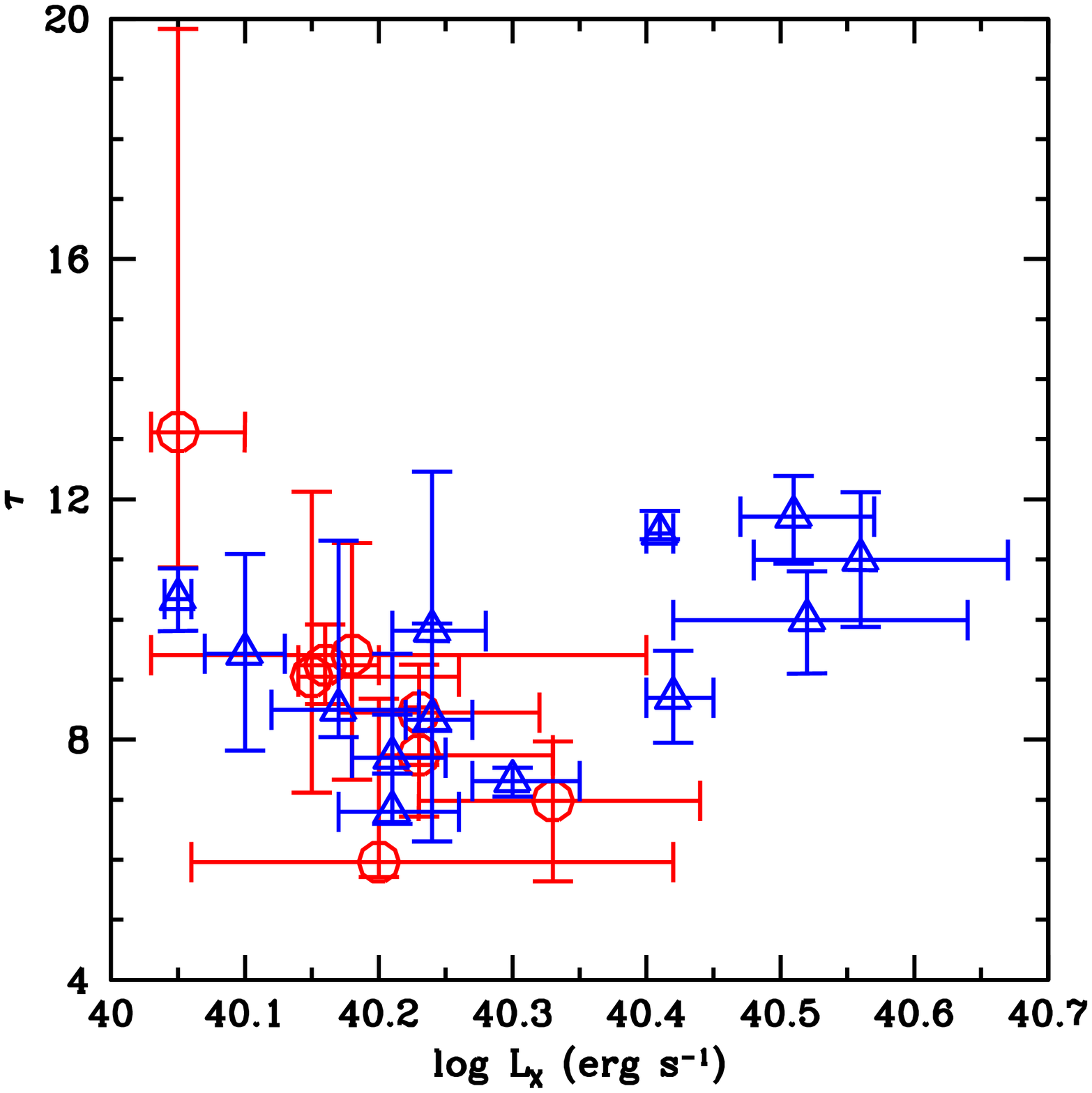}
\includegraphics[width=7.7cm,angle=0]{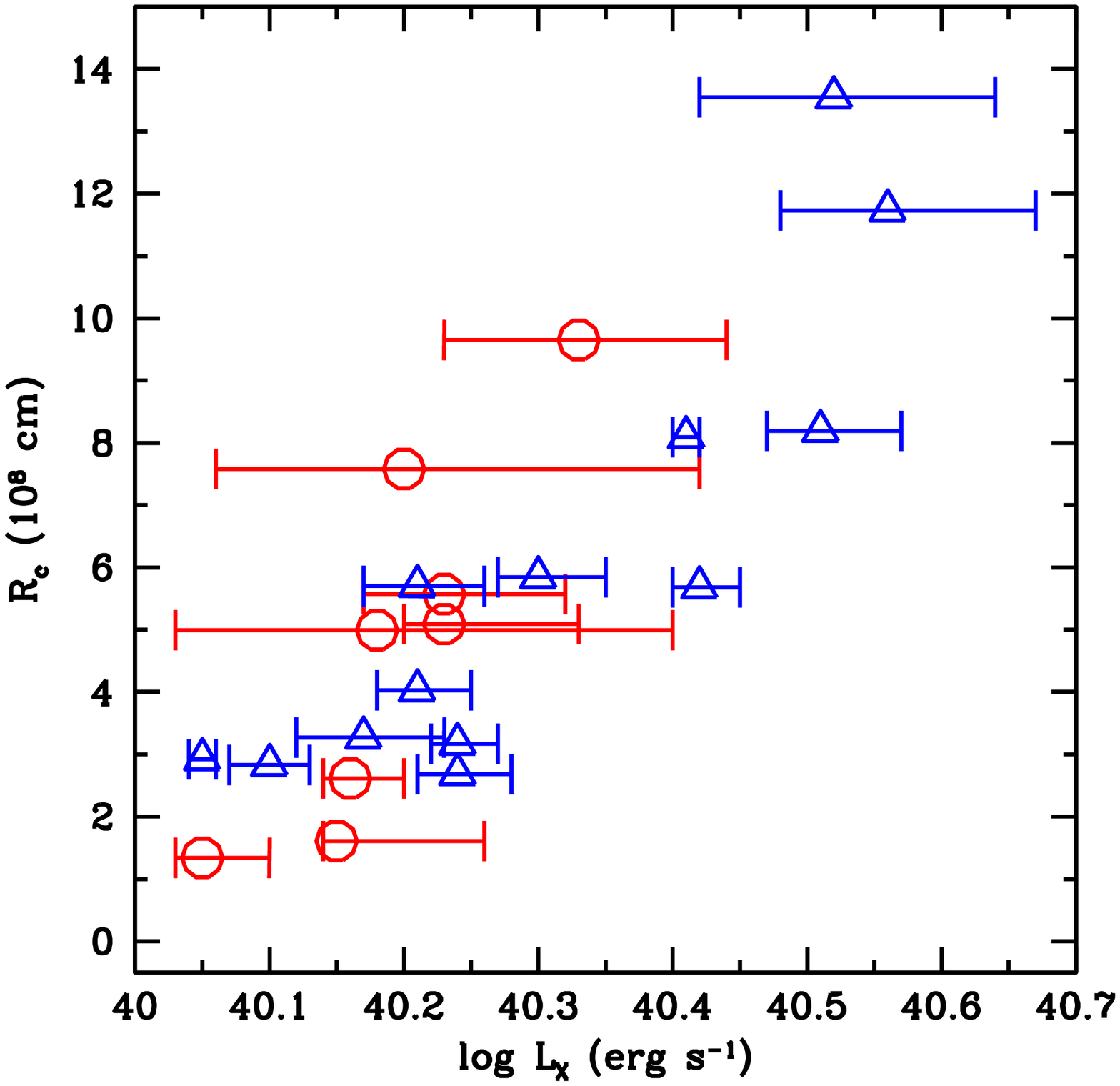}

\caption{{\it Top and middle panels} : The variations of derived parameters from the {\tt nthcomp} model function with $L_{\rm X}$ and $\Gamma$. {\it Bottom panels} : The variations of optical depth and coronal radius with $L_{\rm X}$. The colors and symbols are the same as those used in Figure~\ref{lxVsparams}. See text for more details.}
\label{dev_params}
\end{center}
\end{figure*}

Our base-line model with a black hole suggests that the 
compact object is surrounded by the corona and multi-color 
accretion disc, where the inner region of the disc may be covered by 
the corona. The input seed photons from the accretion disc are injected into 
the corona and the corona Comptonizes these seed photons. Since the source 
exhibits variability, it is important to know how the changes occur 
in the radiation mechanism and geometry of the source. Using the 
{\sc xspec} model function ({\it nthcomp.f}) for the {\tt nthcomp} 
model, we estimated the amplification factor, $A = L_{\rm C}/L_{\rm Input}$, 
where $L_{\rm C}$ and $L_{\rm Input}$ are the luminosities of the 
Comptonizing cloud and the input seed photons, respectively. 
The parameter $L_{\rm Input}$ depends on the luminosity of the disc 
as well as the fraction of input seed photons seen by the corona. 
Moreover, the fraction of input seed photons entering the corona 
depends on the accretion geometry of the system. We derived $A$, 
$L_{\rm C}$, and $L_{\rm Input}$ from the model function for all 
the spectra using the best-fit parameters. The variations of the 
derived parameters with $L_{\rm X}$ are shown in Figure~\ref{dev_params}.

It is clear from the figures that $L_{\rm Input}$ and $L_{\rm C}$
increase with $L_{\rm X}$. Moreover, $L_{\rm Input}$ 
varies more rapidly (by a factor of $\sim 5$) than $L_{\rm C}$ 
(a factor of $\sim 3$) and it appears to be flat 
($\sim 7\times 10^{39}\,\ergsec$) at the highest $L_{\rm X}$. 
The changes in $L_{\rm Input}$ and $L_{\rm C}$ lead 
to the variations of $A$ , which decreases from 6.3 to 3.7 
as $L_{\rm X}$ increases, but then turns to be $\sim 4.5-5$ 
at the highest luminosity (Figure~\ref{dev_params}). 
In the Comptonization context, $\Gamma$ is inversely proportional to $A$. 
We did observe such variability in the case of 
Ho IX X--1, where the source evolves from hard ($\Gamma \sim $ 1.4) 
to soft ($\Gamma \sim $ 1.8) as $A$ decreases.
We also estimated the optical depth $\uptau$ \citep{Zdz98} as,

\begin{eqnarray}
\uptau = \left(\frac{\rm kT_{e}}{\rm m_{e}c^{2}}\right)^{-1/2} \left[\left(\Gamma+\frac{1}{2}\right)^{2} - \frac{9}{4}\right]^{-1/2}\ \ \ 
\end{eqnarray}

The derived $\uptau$ values are consistent with having optically thick 
corona ($\uptau \sim 6-13$) and showed marginal variations with $L_{\rm X}$, 
first dropping and then going up as $L_{\rm X}$ increases. 

\section{Discussion}

We analyzed the \suzaku{} and \xmm{} observations of the bright 
ULX Ho IX X--1 conducted over a period of $\sim 14$ years, to study its 
spectral variability. We systematically studied the spectra with different 
models. The data were better represented by a MCD plus thermal 
Comptonization ({\tt nthcomp}) model. The best-fit spectral parameters 
derived from this model were consistent with the results of broadband 
spectral studies in the 0.3--30 keV energy band using the disc-corona 
plus PL-like tail model \citep[$N_{\rm H} \sim 1.4\times10^{21}~\rm cm^{-2}$, $kT_{\rm in} \sim 
0.3$ keV, $kT_{\rm e} \sim 2.4$ keV;][]{Wal14}. Using the MCD plus thermal Comptonization model, 
we studied the variability behavior of the accretion 
disc plus corona as a function of X-ray luminosity. Our analysis revealed 
that the best-fit model parameters showed strong trends with X-ray luminosity. 
The parameters $N_{\rm H}$ and $\Gamma$ were found to be positively correlated 
with X-ray luminosity, while $kT_{\rm in}$ was negatively correlated. The 
plasma temperature derived from this model 
description ranges $\sim 1-4$ keV, consistent with the low temperature, 
optically thick corona seen in other ULXs \citep{Sto06, Gla09}. Moreover, 
$kT_{\rm e}$ did not show a statistically significant correlation with $L_{\rm X}$, 
although it marginally varied with $L_{\rm X}$. The parameter $\Gamma$ 
significantly varied with $L_{\rm X}$ and exhibited a statistically significant positive 
correlation. It evolved from hard ($\Gamma \sim 1.4$) to soft ($\Gamma \sim 1.8$) 
as $L_{\rm X}$ increases, while at higher X-ray luminosities $\Gamma$ 
turned to be slightly lower ($\Gamma \sim 1.7$). In these observations, the X-ray 
luminosity of Ho IX X--1 varied by a factor of $\sim 3-4$ and the flux 
contribution from the Comptonized component was much higher than that
from the disc component. 

In the MCD plus thermal Comptonization model, one can assume a geometry, 
where a standard cold accretion disc is truncated at radius $R_{\rm tr}$ 
and the inner region contains hot plasma. The plasma Comptonizes the seed 
photons from the outer disc, and the fraction of input seed photons 
that enter the plasma region is related to a solid angle subtending
between the plasma region and the outer disc. For this geometry, 
the input seed photon luminosity must be smaller than the disc luminosity, 
$L_{\rm Input}/L_{\rm Disc} \lesssim 1$, but if $L_{\rm Input} \ll L_{\rm Disc}$, 
it poses an unphysically small solid angle. This geometry has been proposed 
for the ULX NGC 1313 X-1, where the source was at a low flux state \citep{Dew10}. 
For Ho IX X--1, such geometry is consistent with the spectra of {\it XMM}\,5, 
{\it XMM}\,6 and {\it Suzaku}\,7, where $L_{\rm Input} / L_{\rm Disc} \sim 0.4-0.8$. 
The values suggest the solid angle $\Delta \Omega \sim (0.4-0.8)\times 2\pi$. 
The estimated truncation radius from the normalization of the disc component 
for these three spectra ranges between 
$8.7 \times 10^{8} - 2.1\times 10^{9} \rm cm$,
which is corrected for color factor $\kappa=1.7$ (\citealt{Shi95};
the inclination angle of the binary is assumed to be $60^{\circ}$). 
The proposed scenario is consistent with the existence of a 
massive black hole of mass about 50--200\,$M_{\sun}$. 
If we assume a mass of 200\,$M_{\sun}$ for Ho IX X--1, 
the Schwarzschild radius would 
be $r_{\rm s}=2GM/c^{2} \sim 6\times 10^{7}\rm cm$, making the transition 
radius $R_{\rm tr} \sim 15-35\,r_{\rm s}$. Thus, the three spectra were 
consistent with a model where a massive black hole surrounded by a standard 
accretion disc truncated at a radius of $\sim 15-35\,r_{\rm s}$. 

However, majority of the spectra, 18 out of 21, appear to be inconsistent with 
this geometry as $L_{\rm Input} > L_{\rm Disc}$. The observed properties of 
the source instead can be explained by an alternative geometry, 
the ``sandwich model'' \citep{Lia77, Haa93, Sve94}, which has been
successful in describing the observed properties of Galactic 
BHXBs \citep[][and references therein]{Kub04, Don06}. 
In this model, the corona covers the standard accretion disc
and takes some fraction of the total gravitational power, while 
the remaining fraction is dissipated in the disc. The 
corona Comptonizes the seed photons from the underlying disc and a fraction of 
Comptonized photons ($\xi$) impinge on the disc, and get absorbed. 
This geometry is valid only when the fraction $\xi$ is less than 
the maximum value ($\xi < \xi_{\rm max} =1/A$). Such geometry can explain 
the observed properties of NGC 1313 X-1 in the high flux 
state \citep{Dew10}, where the corona covers the entire disc. For Ho IX X--1, 
the MCD component is clearly evident in all the spectra, suggesting
that the corona covers only the inner part of the disc. 
This can be confirmed by estimating the size of the corona region 
$R_{\rm c}$.  Assuming the input seed photon luminosity as a blackbody, 
$\sigma T_{\rm s}^4 2 \pi R_{\rm c}^{2} = L_{\rm Input} $ 
\citep[see also][for more details]{Dew10}, where $T_{\rm s}$ is the temperature 
of the disc. The derived values of $R_{\rm c}$ exhibit an 
increasing trend with $L_{X}$ (see Figure~\ref{dev_params}) and 
$R_{\rm c} \sim 1.4-13.6\times 10^{8}\ \rm cm$, which can definitely 
mask the inner part of the disc. The negative correlation between $kT_{\rm in}$ 
and $L_{\rm X}$ (see Figure~\ref{lxVsparams}) can thus be because of the larger
part of the inner disc covered by the corona.

From our studies, we revealed the existence of a more complex $\Gamma - L_{\rm X}$ 
relation: $\Gamma$ evolves from hard to soft ($\sim 1.4$ to $1.8$) 
as $L_{\rm X}$ increases, while at higher $L_{\rm X}$, $\Gamma$ turns to be
slightly harder ($\sim$ 1.7). The positive $\Gamma - L_{\rm X}$ correlation 
has been reported for Ho IX X--1, based on an absorbed PL fit, with limited 
sample of observations \citep{Kaj09}. The observed values of photon index 
($\Gamma \sim 1.5-2.0$) from the simple PL are also consistent with that from 
our base-line model. We further note that the $\Gamma - L_{\rm X}$ correlation 
has been reported for several other ULXs, for example, NGC 1313 X--1 \citep{Fen06}, 
Antennae X--11 \citep{Fen06a}, NGC 253 X--4, IC 342 X--6, Holmberg II X--1, 
NGC 5204 X--1 and NGC 5408 X--1 \citep{Fen09, Kaj09}, in which their spectra 
were fitted with a PL or a MCD plus PL model. This correlation phase of ULXs 
was explained as an intermediate state with hybrid properties from 
the thermal and steep PL states \citep{Fen09}.

However in the disc-corona model we used, the $\Gamma - L_{\rm X}$ 
correlation can be explained as due to the process of  
the thermal Comptonization of the seed photons from the accretion disc
by the hot corona  \citep[][and references therein]{Zdz03}. 
In this process, the seed photons are variable, and when the seed photon flux 
increases, the X-ray emission becomes softer and stronger 
\citep{Zdz01, Zdz02}. Therefore in this model, the variability of 
$L_{\rm Input}$ should be stronger than that of $L_{\rm C}$. 
Ho IX X--1 is likely such a case. We found that $L_{\rm Input}$ is more 
variable (by a factor of $\sim 5$) than $L_{\rm C}$ and the amplification 
factor decreases with $L_{\rm X}$. 
Moreover, the input seed photon flux for Ho IX X--1 shows the sign 
of saturation at high luminosities. Thus, at high luminosities although 
the radiation flux from the corona has increased, the seed photon flux 
remains more or less same, which should lead to hardening of the spectra 
as is marginally observed. The saturation of the seed photon flux implies 
that while the coronal radiative power has increased, the disc flux remained 
nearly constant or did not increase proportionally. This suggests that at 
high luminosities a larger fraction of the accretion energy is dissipated 
in the corona as compared to disc, perhaps because the corona has become 
larger, covering a greater fraction of the accretion disc.
The variable seed photon flux also affects the coronal parameters. 
Our modeling showed a marginal increase in $kT_{\rm e}$ as $L_{\rm X}$
(or $L_{\rm Input}$) increases, while it drops at the high luminosity. 
The former behavior has been observed in other ULXs, NGC 5204 X--1 
and Holmberg II X--1 \citep{Rob06, Fen09}, while the latter is consistent 
with earlier studies of Ho IX X--1 \citep{Vie10} and IC 342 X--1 \citep{Fen09}. 
The change in $kT_{\rm e}$ is also reflected in the optical depth, which 
first decreases (from $\uptau \sim$ 13 to 6) and then becomes very thick 
($\uptau \sim 12$) at the high luminosity. The variability indicates that 
Ho IX X--1 exhibits a cooler ($kT_{\rm e} \sim 2$ keV) and very thick corona 
($\uptau \sim 12$) at different luminosity values. We note that in NGC 1313 
X--2, the different corona states (`very-thick' and `thick') correlate with 
luminosity and the source becomes more luminous in the `very-thick' state \citep{Pin12}.
The comparison indicates the more complex variability behavior in Ho IX X--1. 

As mentioned above, the source turns to be marginally hard at the high 
luminosities, i.e., $L_{\rm X} > 2\times10^{40}\,\ergsec$. 
This turn-over of $\Gamma$ in the $\Gamma - L_{\rm X}$ plane is significant,
and results a change in the slope of the correlation. Interestingly, the optical 
depth also exhibits similar variability pattern as the source appears to be 
optically very thick ($\uptau \sim 12$) above the luminosity $2\times10^{40}\,\ergsec$. 
In \citet{Pin14}, the softening behavior was observed for Ho IX X--1 in the 
`thick' state as the intensity increases in the hardness-intensity diagram
\citep[left panel of Figure 6 in][]{Pin14}. We also observe such softening 
variability behavior below the luminosity $2\times10^{40}\,\ergsec$ 
(mostly in the `thick' state) in the $\Gamma - L_{\rm X}$ plane.
However, the behavior of low-luminosity sources \citep[NGC 253 X--1 and 
NGC 1313 X--2 in the `thick' state;][]{Pin14} is different, where these 
sources become marginally hard as the intensity increases to a level consistent with 
the `very thick' state seen in a few ULXs. Then in the `very-thick' state, NGC 1313 X--2 
turns to have the softening behavior as the intensity increases. Thus, comparing 
Ho IX X--1 to NGC 1313 X--2, these sources have different behavior below and above 
a certain luminosity threshold. The threshold is also different in the two sources, 
as for the former, its value is $\sim 2\times10^{40}\,\ergsec$, while it can be 
approximated to be $5\times10^{39}\,\ergsec$ \citep[see Figure 3 in][]{Pin12} for 
the latter.

In the models with a corona, the energy balance is achieved by adjusting the 
coronal parameters: varying either the electron temperature alone or both 
$kT_{\rm e}$ and $\uptau$. For example, if the optical depth is constant, the 
corona adjusts the electron temperature itself to the variable seed photon 
flux. Thus, the energy balance is satisfied by an increased cooling of the plasma ($kT_{\rm e}$ 
decreases), which results in the increase of $\Gamma$ and softer spectrum 
\citep{Zdz01, Zdz02, Zdz03}. However, it is also possible 
that the variation in the photon index is due to the optical depth variation 
\citep[][and reference therein]{Sve94a,Haa97}. In such cases, the energy balance is 
satisfied by adjusting $\uptau$ and the softening of the spectrum is associated 
with an increase of $kT_{\rm e}$ \citep{Nic00, Chi02, Zdz03}. The latter explanation 
seems to be consistent with observed behavior of the coronal parameters of 
Ho IX X--1 below a luminosity of $2\times10^{40}\,\ergsec$. 
At the luminosity above $2\times10^{40}\,\ergsec$, the reason for the change 
in the slope of the $\Gamma - L_{\rm X}$ relation appears to be the saturation 
of the input seed photon luminosity. Considering this and the energy balance 
of the system in the high luminosity, one would expect that at constant 
optical depth the coronal electron temperature increases as $L_{\rm X}$ 
increases, while we observed an opposite behavior. Thus, this behavior 
can be explained as when $L_{\rm X}$ increases above $2\times10^{40}\,\ergsec$, 
the corona becomes not only more energetic but also more mass loaded, 
which increases the coronal density and the optical depth of the source. 
In such a case, the decrease in the mean energy per scattering is largely 
compensated by an increase of the average number of scatterings, which 
produces the observed hardening of the spectrum. 

While in this work, we have interpreted Ho IX X--1 as a massive black hole 
accreting at near-Eddington rate, the source has also been interpreted as a 
stellar-mass black hole accreting at super-critical rates leading to a strong 
radiation driven winds \citep{Sut13, Pin14, Lua16}. In this scenario, the wind 
may be sufficiently dense and cover a large fraction of the outer disc. When 
a ULX system is observed face-on, the inner region around a black hole emits 
a hard spectrum, and at high inclination angles, the line-of-sight passes 
through the outer, cold region of the wind and a softer spectrum is 
observed \citep[see e.g.][]{Mid11b, Sut13, Mid14}. 
\cite{Sut13} studied the spectral variability of a large sample of ULXs, 
using a doubly absorbed MCD plus PL model, and empirically classified the ULX 
population into three spectral regimes, referred as {\it broadened-disc}, 
{\it hard ultraluminous}, and {\it soft ultraluminous}. These spectral states 
can be explained as being due to different viewing angles to the sources. 
However, Ho IX X--1 shows spectral characteristics of the {\it hard ultraluminous} 
and {\it broadened-disc} class at different times \citep{Sut13}. This was 
confirmed by another detailed study of large sample of ULXs, where 
color-color and hardness-intensity diagrams were used \citep{Pin14}. 
\cite{Lua16} also characterized the spectral evolution of Ho IX X--1 
below 10 keV. They used a two-thermal component model, {\tt diskbb}+{\tt comptt}, 
similar to what has been used in this work and our results are consistent 
with theirs. For example, the flux variations primarily came from the 
hard component and at the high-luminosity end, the spectra of the hard 
component is harder. The difference between theirs and this work is in 
the interpretation of the spectral variability, where they consider the 
wind model of a stellar-mass black hole, while here we consider the 
system to be a massive black hole accreting at near-Eddington rates. 
We note that no direct evidence for a strong disc wind has been found 
in Ho IX X--1 \citep{Wal12, Wal13}, although some residuals in 
the source's soft spectra have been suggested to be line features from 
the wind \citep{Mid15}. In any case, our study has revealed 
interesting variability features of the source, in particular the 
$\Gamma - L_{X}$ correlation. These features can be considered 
as significant constraints, to be explained by any scenario proposed 
for this ULX.

\section*{Acknowledgements}

We thank the anonymous referee for the constructive 
comments and suggestions that improved this manuscript. 
This research has made use of data obtained from the High Energy 
Astrophysics Science Archive Research Center (HEASARC), provided 
by NASA's Goddard Space Flight Center. This research was supported 
by the National Program on Key Research and Development Project 
(Grant No. 2016YFA0400804) and the National Natural Science Foundation 
of China (11373055, 11633007). VJ acknowledges the IUCAA Visitor's 
Program and the financial support from Chinese Academy of Sciences 
through President's International Fellowship Initiative (CAS PIFI, 
Grant No. 2015PM059). Z.W. acknowledges the support by the CAS/SAFEA 
International Partnership Program for Creative Research Teams.

\software{{\sc heasoft} (v 6.15.1), {\sc sas} (v 14.0), {\sc xspec} (v 12.8.1g; \citealt{Arn96})}


\end{document}